\documentclass[%
 aps,pre,
 amsmath,amssymb,
reprint,%
]{revtex4-2}

\usepackage{graphicx}% Include figure files
\usepackage{dcolumn}% Align table columns on decimal point
\usepackage{bm}% bold math
\usepackage{amsfonts,amsthm,mathbbol}
\usepackage[utf8]{inputenc}
\usepackage[T1]{fontenc}
\usepackage{mathptmx}
\usepackage{color}

\begin{document}

\title[The effective diffusion constant of stochastic processes with spatially periodic noise]{The effective diffusion constant of stochastic processes with spatially periodic noise}

\author{Stefano Giordano}%
\email{stefano.giordano@univ-lille.fr}  
\affiliation{University of Lille, CNRS, Centrale Lille, Univ. Polytechnique Hauts-de-France, UMR 8520 - IEMN - Institut d'{\'E}lectronique, de Micro{\'e}lectronique et de Nanotechnologie, F-59000 Lille, France
}

\author{Ralf Blossey}
\email{ralf.blossey@univ-lille.fr}
\affiliation{University of Lille, Unit{\'e} de Glycobiologie Structurale et Fonctionnelle (UGSF), CNRS UMR8576, F-59000 Lille, France}

\date{\today}% It is always \today, today,
             %  but any date may be explicitly specified

%%%%%%%%%%%%%%%%%%%%%%%%%%%%%%%%%%%%%%%%%%%%%%%%%%%%%%%%%%%%%%%%%%%%%%%%%%%%%%%%
\begin{abstract}
We discuss the effective diffusion constant $D_{{\it eff}}$ for stochastic processes with spatially-dependent noise. 
Starting from a stochastic process given by a Langevin equation, different drift-diffusion equations can be derived depending on the choice of the discretization rule $ 0 \leq \alpha \leq 1$. 
We initially study the case of periodic heterogeneous diffusion without drift and we determine a general result for the effective diffusion coefficient $D_{{\it eff}}$,  which is valid for any value of $\alpha$. 
We study the case of periodic sinusoidal diffusion in detail and we find a relationship with Legendre functions.  
Then, we derive $D_{{\it eff}}$ for general $\alpha$ in the case of diffusion with periodic spatial noise and in the presence of a drift term, generalizing the Lifson-Jackson theorem. 
Our results are illustrated by analytical and numerical calculations on generic periodic choices for drift and diffusion terms.
\end{abstract}

\maketitle

%%%%%%%%%%%%%%%%%%%%%%%%%%%%%%%%%%%%%%%%%%%%%%%%%%%%%%%%%%%%%%%%%%%%%%%%%%%%%%%%

\section{Introduction}

The concept of heterogeneous diffusion with a spatially varying diffusivity is widely discussed in the literature to describe anomalous diffusion processes, see, e.g.,  \cite{silva2011,cherstvy2013,metzler2014,cherstvy2017,wang2020,ritschel2021,vinod2022,ribeiro2023}. Multiplicative noise plays an essential role in describing the behavior of several physical and biological phenomena including the transmission of signals in neuron models \cite{bauermann2019,zhu2021}, phenotypic variability and gene expression \cite{liu2004,frigola2012}, the stochastic thermodynamics of holonomic systems \cite{bianco2001,manca2016,giordano2019}, the ballistic-to-diffusive transition of heat propagation \cite{landi2014,palla2020}, 
the fluctuations effects in lasers and semiconductors \cite{landauer1962,haken1975,haenggi1982}, the statistical theory of turbulence \cite{fuchs2022,birnir2013}, and the modeling of stock prices, particularly through the  Black-Scholes model of option pricing \cite{hull2021,bouchaud2009}. This long but clearly non-exhaustive list of examples shows that heterogeneous diffusion is an extremely useful and versatile tool for the comprehension of a large variety of phenomena ranging from physics to biology, and to finance.

Recently, we have addressed the role of the discretization rule of stochastic processes with heterogeneous, i.e. spatially-dependent noise, in both long- 
and short-time limits \cite{giordano2023,dupont2024}. The discretization rule refers to the rule of integration of the Langevin equation, and commonly involves the introduction
of a real parameter $\alpha$, with $0 \leq \alpha \leq 1$, with the common cases being the Fisk-Stratonovich (midpoint) rule, $\alpha = 1/2$ \cite{fisk1963,stratonovich1966}, or the endpoint rules $\alpha = 0$ (It\^o) \cite{ito1950} and $\alpha = 1$ (H\"anggi-Klimontovich) \cite{haenggi1982,klimontovich1995}. 
In our earlier work \cite{giordano2023,dupont2024}, we established a number of limitations for the 
existence of the probability distributions in short-and long-time regimes of certain Fokker-Planck equations, thereby demonstrating the relevance of
the discretization rule for the physical context in which the corresponding equations might be employed.

In this work, we continue our discussion of the importance of the discretization rule. Here, we address the derivation of the effective diffusion constant for stochastic heterogeneous (drift-)diffusion processes. 
Effective diffusion constants are well established quantities 
describing the long-time and large length scale transport properties in heterogeneous systems, see, among many others, Refs. \cite{jackson1963,weissberg1963,crank1975,mccarty1988,revathi1993,drexler2023,defaveri2023,spiechowicz2023}. 
They characterize, e.g., the mean-squared displacement (MSD) of particles moving in an external potential at late times via \cite{sorkin2023}
\begin{equation}
    \langle [X(t) - X(0)]^2\rangle \simeq 2D_{{\it eff}}t\, .
\end{equation}
The effective diffusion constant $D_{{\it eff}}$ can be determined for drift-diffusion equations, i.e. the Fokker-Planck equation, or, more simply diffusion laws in heterogeneous media, in which, at the level of the Langevin equation, the discretization rule of the stochastic process in general matters, in contrast to the process occurring under the action of simple (constant) noise \cite{risken1989,oksendal2003,coffey2004,gardiner2009}. 
A general formula for the effective diffusion constant was provided already long ago by 
Lifson and Jackson \cite{lifson1962} and also others
(see below); however, to the best of our knowledge, without addressing the role of the discretization of the underlying stochastic process.

{
While most studies on heterogeneous diffusion consider arbitrary spatially-dependent noise $g(x)$, in this paper we focus our attention to the case of {\it spatially periodic} noise $g(x)$. }
Spatially periodic 
noise has been discussed in the literature early on,
see the classic work by B\"uttiker \cite{buettiker1987} and Landauer \cite{landauer1988}.
B\"uttiker specifically considered the interplay between a periodic drift and a periodic noise, and observed that current flow can arise when the two modulations are out of phase \cite{buettiker1987}.
Landauer studied systems with nonuniform temperature observing that particles  move out of the hot
regions with greater velocity than out of the cold regions, eventually modifying the overall dynamics \cite{landauer1988}.
Recently, the problem has been taken up again by us in the context of the motion of the Brownian particle in a {tilted periodic potential (also called washboard potential)} \cite{breoni2022} - a celebrated problem with a huge literature, see the references listed in Ref. \cite{breoni2022}.

Our focus in the present paper is the determination of the effective diffusion coefficient with an arbitrary value of $0 \leq \alpha \leq 1$. We will first address the case of the absence of drift, and then combining drift and diffusion. Thereby we obtain a generalization the Lifson-Jackson formula for general $\alpha$, a result we believe has not been obtained previously.

The structure of our paper is as follows. Section \ref{hetdiff}
presents the mathematical formulation of the problem studied in
this paper {and outlines the logic followed in our developments}. In Section \ref{secstrato}, we first calculate the effective diffusion constant $D_{\it eff}$ in the Fisk-Stratonovich case with $\alpha=1/2$. Here, the determination of $D_{\it eff}$ can be performed analytically, for which we present two slightly alternative versions {(the second one in Appendix \ref{appeA})}. Our result is then applied to an exemplary case, where the heterogeneous diffusion varies sinusoidally in space.
In this case we determine not only $D_{\it eff}$  but also the full probability density of the process.
In Section \ref{general}, we propose a method to obtain the effective diffusion constant for the periodic heterogeneous diffusion problem with arbitrary stochastic interpretation, i.e. for any value of $\alpha$. 
Finally, in Section \ref{lija}, we consider an additive drift term in the Langevin equation to study the combined effects of periodic drift and periodic diffusion. We prove a generalized form of the Lifson-Jackson theorem, which allows us to obtain the effective diffusion coefficient under these conditions. 
As a special case, we numerically study the problem when  drift and diffusion are both sinusoidal, with an arbitrary phase shift between the two terms. We then discuss the physical phenomena induced by diffusion-drift interaction in this particular case.

\section{Heterogeneous diffusions: model definition and the calculation of $D_{\it eff}$}

\label{hetdiff}

We start our discussion with the case of heterogeneous particle diffusion problems in the absence of forces, i.e. we consider stochastic differential equations of Langevin type
\begin{align}
\label{lgstoc}
    \frac{\mathrm{d} x}{\mathrm{d} t}=g(x)\xi(t),
\end{align}
where $g(x)$ is a spatially-dependent multiplicative noise with $g(x)> 0$ for any $x\in\mathbb{R}$.
The process $\xi(t)$ in Eq. (\ref{lgstoc}) is a Gaussian white noise with average value $\mathbb{E}(\xi(t))=0$, and correlation $\mathbb{E}(\xi(t)\xi(\tau))=2\delta(t-\tau)$, where $\delta(t)$ is the  Dirac delta function. 

In order to properly define the mathematical meaning of Eq. (\ref{lgstoc}), we need to specify the type of stochastic interpretation adopted, or equivalently the discretization parameter $\alpha$, as discussed in the
Introduction.
The equivalent diffusion or Fokker-Planck equation associated with Eq. (\ref{lgstoc}) contains $\alpha$ explicitly as a simple parameter. For general $ \alpha $, it is given by the expression
{
\begin{align}
    \frac{\partial W(x,t)}{\partial t}=\frac{\partial}{\partial x}\left\lbrace g^{2\alpha}(x)\frac{\partial }{\partial x}\left[g^{2(1-\alpha)}(x)W(x,t)\right]\right\rbrace,
    \label{fpgenintro}
\end{align}
}
for the probability density $W(x,t)$ \cite{giordano2023}.
For the above mentioned cases of $\alpha =  1/2, 1,0$, one thus obtains three distinct expressions of
Fokker-Planck type
 \begin{align}
 \label{strato}
        \frac{\partial W}{\partial t}&=\frac{\partial}{\partial x}\left[g\frac{\partial}{\partial x}\left(gW\right)\right],&\alpha=1/2,\\
        \label{antiito}
        \frac{\partial W}{\partial t}&=\frac{\partial}{\partial x}\left[g^2\frac{\partial W}{\partial x}\right],&\alpha=1,\\
        \label{ito}
        \frac{\partial W}{\partial t}&=\frac{\partial^2}{\partial x^2}\left[g^2W\right],&\alpha=0\, 
\end{align}
that have all appeared before in the scientific literature. The first equation is referred to as Wereide's equation \cite{wereide1914}, the second corresponds to the classical Fick law \cite{fick1855}, and the third is known as Chapman's law \cite{chapman1928}. 

Wereide's diffusion law has been originally obtained to study the particles diffusion in a region where there is a spatially varying  temperature field \cite{wereide1914}. 
Today, the Wereide law has been shown to correctly describe biological processes of invasion into periodically fragmented environments 
\cite{shigesada2003}. 
Fick's law, firstly introduced by Adolf Fick in 1855 \cite{fick1855}, governs the transport of mass through diffusive phenomena, as largely confirmed by  experimental results. Fick's law is in strong analogy with other mathematical expressions describing similar phenomena: the Darcy law for the hydraulic flow in porous media, the microscopic Ohm law describing the charge transport in electrical conductive materials, and Fourier's law explaining the heat transport in thermally conductive media. 
Finally, Chapman's law has been introduced, by means of statistical mechanics arguments, to describe diffusion processes in non-uniform fluids \cite{chapman1928}.
Recently, it has been demonstrated that Chapman's diffusion law describes protein transport in heterogeneous biological environments better than the Fick diffusion law \cite{chae2024}. 
While the comparison among the different stochastic interpretations is performed in Refs.\cite{kampen1981,sokolov2010}, the corresponding diffusion processes are analyzed and discussed in Refs.\cite{upadhyay2021,kim2021,alfaro2022}.

On the level of the probability density $W(x,t)$, 
we now want to define a homogeneous diffusion equation 
\begin{equation}
\frac{\partial W}{\partial t} = D_{\it eff} \frac{\partial^2 W}{\partial x^2}
\end{equation}
that should represent, in a sense that needs to be precisely specified, the homogenized version of the previous Fokker-Planck equations. Thus arises the question on how
to define the effective diffusion constant. One possibility is \cite{weaver1979}
\begin{equation}
    D_{\it eff} \equiv \overline{D}=\lim_{t\to \infty}\frac{\left\langle x^2 \right\rangle}{2t}\, .
    \label{firsteff}
\end{equation}
Another possible definition of the effective diffusion constant is the given by the expression
\begin{align}
    D_{\it eff} \equiv \frac{a}{2\overline{T}_{FP}},
    \label{secondeff}
\end{align}
where $a$ is the length of a finite box and $\overline{T}_{FP}$ is the mean first passage time. 
Both expressions are equivalent \cite{weaver1979}, 
and we will indeed apply them both.

{In the following, we are concerned with determining the effective diffusion coefficient in the case of a periodic function $g(x)$, by using Eqs.(\ref{firsteff}) and (\ref{secondeff}). 
The proposed approach will be valid even with certain types of discontinuity in the function $g(x)$, as long as it remains positive and finite. For example, we can admit a finite discontinuity in the derivative to represent sawtooth functions, or we can admit a finite discontinuity in the function itself to describe rectangular waveforms. More generally, we can say that the theory is valid for periodic functions $g(x)$ that can be developed in Fourier series.

In the following Section \ref{secstrato}, we will first determine the effective diffusion coefficient in the case of periodic heterogeneous diffusion studied under the hypothesis of the Fisk-Stratonovich interpretation. 
Indeed, in this case the general solution of Eq. (\ref{strato}) is known for an arbitrary function $g(x)$ and, of course, it can be used for a periodic $g(x)$. The knowledge of the probability density allows the application of Eq. (\ref{firsteff}) since we can directly calculate the average value $\left\langle x^2 \right\rangle$, and eventually obtain the effective diffusion constant  under the Fisk-Stratonovich interpretation. For this reason, we first started to approach the problem when $\alpha=1/2$. 
Unfortunately, for the other cases with $\alpha\neq 1/2$, the corresponding Fokker-Planck equations cannot be solved in closed form in order to obtain the relevant probability density. That is why, in the absence of the probability density, we preferred Eq. (\ref{secondeff}) to determine the effective diffusion coefficient.  
Even before carrying out this calculation, note that in the case of the H\"anggi-Klimontovich stochastic interpretation, we obtain the classical Fick diffusion law described by Eq. (\ref{antiito}).
In this case, the effective diffusion constant is well-known thanks to several homogenization methods applied to one-dimensional or stratified media and is given by  $D_{\it eff}=\left\langle{1}/{g^2} \right\rangle^{-1}$  \cite{milton2004,kim2011,camacho2013,giordano2014a,giordano2014b}. 
Here, $\left\langle \cdot \right\rangle$ represents the mean value of the argument over one period. We observe that this expression is different from the one obtained for the case with $\alpha=1/2$, which is $D_{\it eff}={\left\langle{1}/{g} \right\rangle^{-2}}$ (see next Section \ref{secstrato}). This proves that the discretization parameter plays an important role in the characterization of effective diffusion. 
This fact motivated us to look for a general result, based on Eq. (\ref{secondeff}), and we demonstrate in Section \ref{general} that 
$D_{\it eff} = \left\langle {1}/{g^{2\alpha}} \right\rangle^{-1}\left\langle {1}/{g^{2-2\alpha}}\right\rangle^{-1}$. This result is consistent with previous particular cases and generalizes the determination of the effective diffusion coefficient to any value of $\alpha$. It will be studied in detail for the case of a sinusoidal function $g(x)$. To conclude, in Section \ref{lija}, we propose a generalization of the classical Lifson-Jackson theorem for the case with spatially periodic drift superposed to a periodic heterogeneous diffusion. This problem is approached by means of a formal analogy with the case without drift, solved in the previous Sections. 
}

\section{Periodic heterogeneous diffusion in the Fisk-Stratonovich interpretation}

\label{secstrato}

{We determine here the effective diffusion coefficient for a periodic heterogeneous diffusion model under the Fisk-Stratonovich interpretation.
Our starting point for this calculation is the closed form expression of the propagator of Eq.(\ref{strato}) that we recently formulated ($\alpha=1/2$) \cite{dupont2024}.} It is given by
\begin{align}
    W(x,t;x_0,t_0)=\frac{\exp\left[-\frac{1}{4(t-t_0)}\left(\int_{x_0}^x\frac{\mathrm{d}\eta}{g(\eta)}\right)^2\right]}{g(x)\sqrt{4\pi(t-t_0)}},
    \label{propar}
\end{align}
for the deterministic initial condition $W(x,t_0;x_0,t_0)=\delta(x-x_0)$.
The Wiener process is obviously retrieved when $g(x)$ is a constant. {The result in Eq. (\ref{propar}) is correct for any function $g(x)$, and in particular is valid for a periodic function $g(x)=g(x+L)$ with period $L\in\mathbb{R}$. The periodicity assumption will always be adopted in the following.} Eq. (\ref{propar}) can be specialized for $x_0=0$ and $t_0=0$, yielding the probability density $\rho(x,t)=W(x,t;0,0)$. 
In this particular case, it is interesting to study the effective diffusion constant, as defined in Eq. (\ref{firsteff}). We can write
\begin{eqnarray}
\nonumber
    \overline{D}&=&\lim_{t\to \infty}\frac{\left\langle x^2 \right\rangle }{2t}=\lim_{t\to \infty}\frac{1}{2t}\int_{-\infty}^{+\infty}x^2\rho(x,t)\mathrm{d} x\\
        &=&\lim_{t\to \infty}\frac{1}{2t}\int_{-\infty}^{+\infty}x^2\frac{\exp\left[-\frac{1}{4t}\left(\int_0^x\frac{\mathrm{d}\eta}{g(\eta)}\right)^2\right]}{g(x)\sqrt{4\pi t}}\mathrm{d} x.
    \label{dave}
\end{eqnarray}
If $g(x)$ is periodic, bounded and strictly positive, $1/g(x)$ is also periodic, bounded, and strictly positive and we can use the Fourier series representation
\begin{align}
    \frac{1}{g(x)}=\sum_{k=-\infty}^{+\infty}C_k\exp{\left(\frac{2\pi i k x}{L}\right)},
\end{align}
where 
\begin{align}
    C_k=\frac{1}{L}\int_{0}^{L}\frac{1}{g(x)}\exp{\left(-\frac{2\pi i k x}{L}\right)}\mathrm{d} x,
\end{align}
with $C_{-k}=C_k^*$, since $g(x)\in\mathbb{R}$ (in particular, it means that $C_0$ is real).
We define $\mathcal{D}(x)=\int_0^x\frac{\mathrm{d}\eta}{g(\eta)}$, and we easily get
\begin{align}
\label{dcp}
    \mathcal{D}(x)&=\sum_{k=-\infty}^{+\infty}C_k\int_{0}^{x}\exp{\left(\frac{2\pi i k x}{L}\right)}\mathrm{d} x\\
    \nonumber
    &=C_0 x +\sum_{\begin{subarray}{c}
			k=-\infty\\
			k\neq 0
		\end{subarray}}^{+\infty}\frac{C_k L}{2\pi i k}\left[\exp{\left(\frac{2\pi i k x}{L}\right)}-1\right]\\
    \nonumber
    &=C_0 x +p(x),
\end{align}
where
\begin{align}
    p(x)=\sum_{\begin{subarray}{c}
			k=-\infty\\
			k\neq 0
		\end{subarray}}^{+\infty}\frac{C_k L}{2\pi i k}\left[\exp{\left(\frac{2\pi i k x}{L}\right)}-1\right]
    \label{pdx}
\end{align}
is a periodic bounded function.
Since $\mathcal{D}'(x)=1/g(x)$, we can rewrite the effective diffusion constant as
\begin{eqnarray}
    \overline{D}=\lim_{t\to \infty}\frac{1}{2t}\int_{-\infty}^{+\infty}x^2\frac{\exp\left[-\frac{\mathcal{D}^2(x)}{4t}\right]}{\sqrt{4\pi t}}\mathcal{D}'(x)\mathrm{d} x,
    \label{dave1}
\end{eqnarray}
where
{
\begin{align}
    \mathcal{D}^2(x)=\left[C_0x+p(x)\right]^2=x^2\left[C_0^2+2C_0\frac{p(x)}{x}+\frac{p^2(x)}{x^2}\right].
\end{align}
}
When the time $t$ is large, the exponential in Eq. (\ref{dave1}) is increasingly flat and close to one, and therefore, the area for large values of $x$ becomes more and more important in calculating the integral. Moreover, $p(x)$ is bounded and periodic and therefore $p(x)/x\to 0$ when $x\to \pm \infty$. Hence, we can write
\begin{eqnarray}
\nonumber
    \overline{D}=\lim_{t\to \infty}\frac{1}{2t}\int_{-\infty}^{+\infty}x^2\frac{\exp\left(-\frac{C_0^2 x^2}{4t}\right)}{\sqrt{4\pi t}}\sum_{k=-\infty}^{+\infty}C_k\exp{\left(\frac{2\pi i k x}{L}\right)}\mathrm{d} x,\\
    \label{dave2}
\end{eqnarray}
and the integral can be calculated using
\begin{eqnarray}
\int_{-\infty}^{+\infty}x^2e^{-ax^2}e^{ibx}dx=\frac{\sqrt{\pi}}{a^{3/2}}e^{-\frac{1}{4}\frac{b^2}{a}}\left(\frac{1}{2}-\frac{b^2}{4a}\right).
\label{int1}
\end{eqnarray}
The application of Eq. (\ref{int1}) to each term of Eq. (\ref{dave2}), for $k$ ranging from $-\infty$ to $+\infty$, shows that only the term for $k=0$ contributes to the result, which is eventually obtained in the simple form 
\begin{align}
    \overline{D}=\frac{1}{C_0^2}={\left\langle\frac{1}{g} \right\rangle^{-2}},
    \label{deff}
\end{align}
where the operator $\left\langle \cdot \right\rangle$ represents the mean value of the argument over one period. 
{This expression can be further confirmed by an alternative derivation discussed for the sake of completeness in Appendix \ref{appeA}.}

\begin{figure*}[t!]
\includegraphics[width=8.5cm]{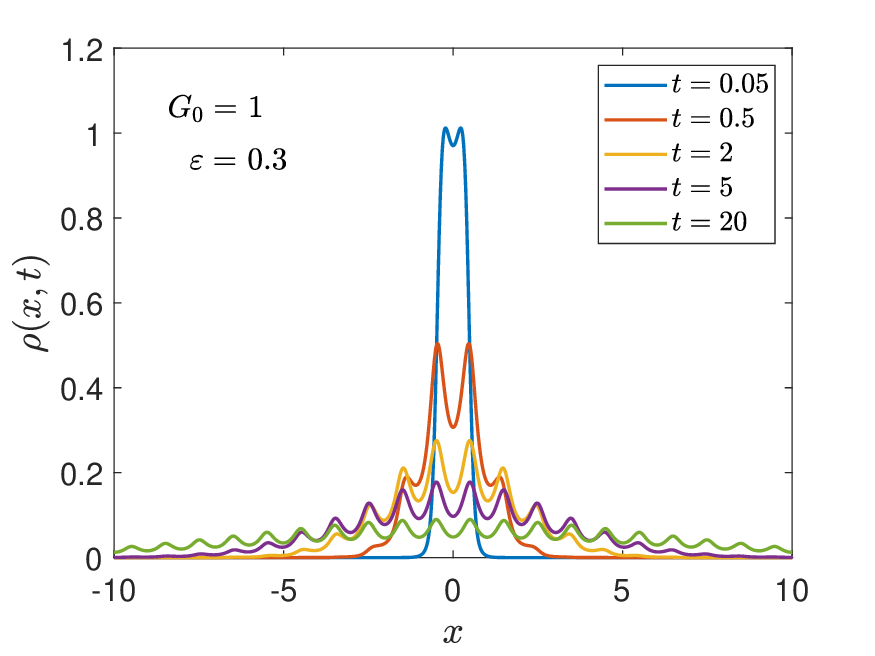}
\includegraphics[width=8.5cm]{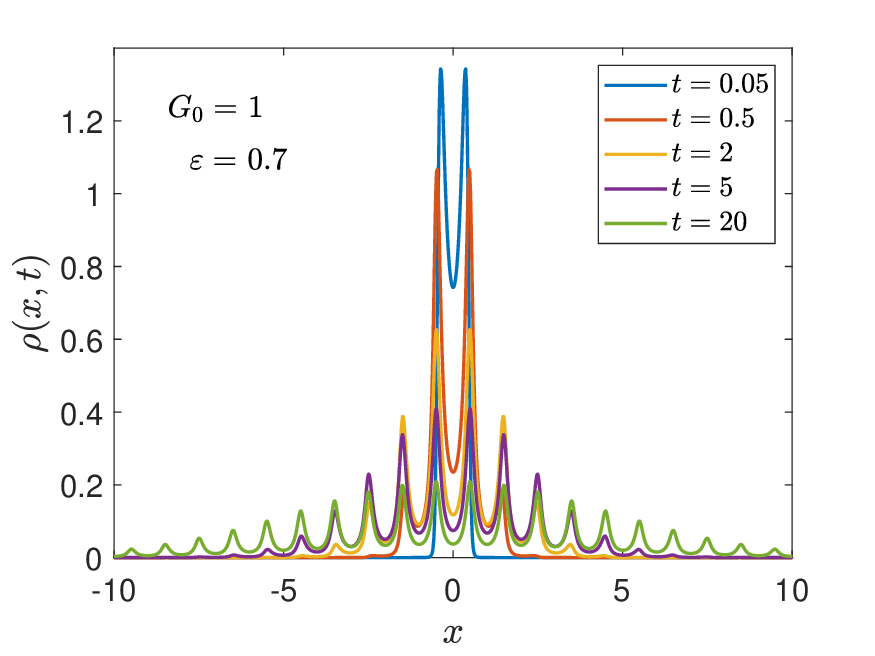}
\includegraphics[width=8.5cm]{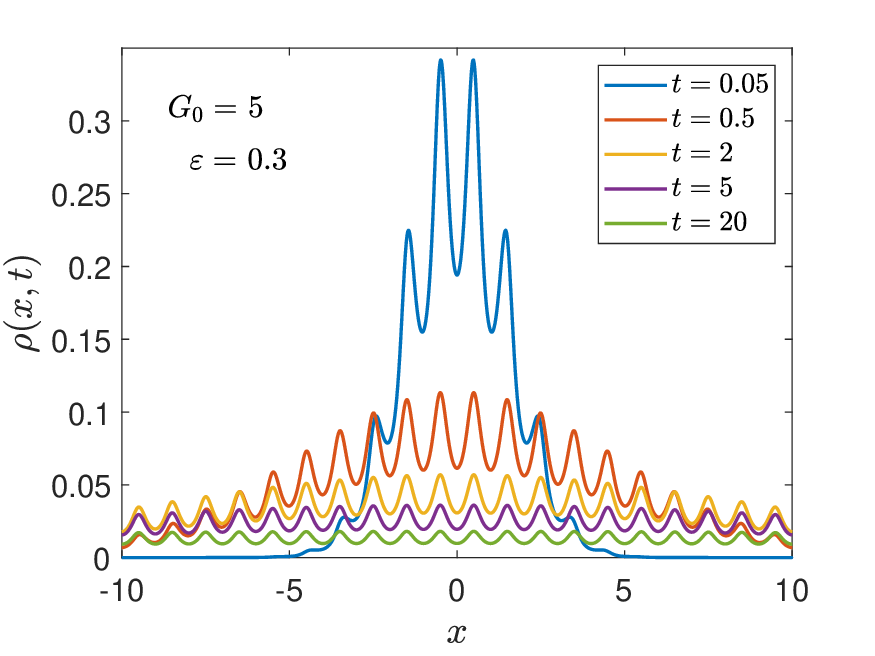}
\includegraphics[width=8.5cm]{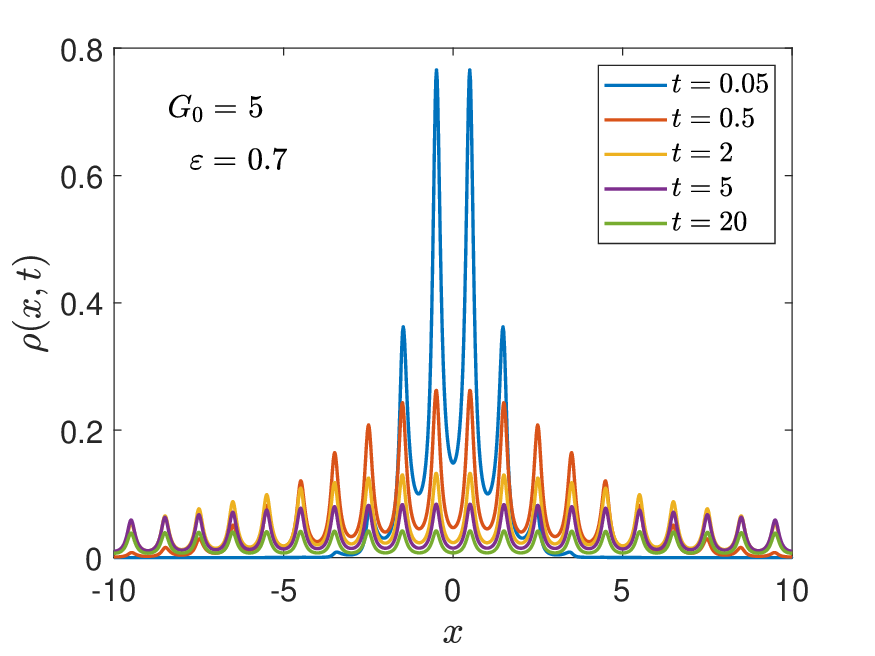}
\caption{\label{dens} Evolution of the probability density for a system with heterogeneous diffusion described by the sinusoidal behavior in Eq. (\ref{trig}). We represented the density $\rho(x,t)$ as a function of $x$, and parameterised by $t$. We considered four cases with the following values of $G_0$ and $\varepsilon$: (i) $G_0=1$ and $\varepsilon=0.3$; (ii) $G_0=1$ and $\varepsilon=0.7$; (iii) $G_0=5$ and $\varepsilon=0.3$; (iv) $G_0=5$ and $\varepsilon=0.7$. We always considered $L=1$. }
\end{figure*}

We now consider the simplest case of periodic diffusion, which is described by a sinusoidal profile
\begin{align}
    g(x)=G_0\left(1+\varepsilon\cos\frac{2\pi x}{L}\right),
    \label{trig}
\end{align}
where $G_0>0$ and $\varepsilon^2<1$. These limitations ensure that we always have $g(x)>0$ $\forall x\,\,  \in\mathbb{R}$.
The Fourier coefficients can be calculated as 
\begin{align}
\nonumber
    C_k&=\frac{1}{L}\int_{0}^{L}\frac{1}{G_0\left(1+\varepsilon\cos\frac{2\pi x}{L}\right)}\exp{\left(-\frac{2\pi i k x}{L}\right)}\mathrm{d} x\\
        &=\frac{1}{\pi G_0}\int_{0}^{\pi}\frac{\cos (ky)}{1+\varepsilon\cos y}\mathrm{d} y,
\end{align}
where we used the change of variable $y=2\pi x/L$ and the parity of the cosine function (here $k\in\mathbb{Z}$).
By considering the integral \cite{gradshteyn1965,abramowitz1970}
\begin{align}
\int_{0}^{\pi}\frac{\cos (ny)}{1+\varepsilon\cos y}dy=\frac{\pi}{\sqrt{1-\varepsilon^2}}\left(\frac{\sqrt{1-\varepsilon^2}-1}{\varepsilon}\right)^{n},
\end{align}
valid for $n\in\mathbb{N}$ and $\varepsilon^2<1$, we directly obtain
\begin{align}
C_k=\frac{1}{G_0\sqrt{1-\varepsilon^2}}\left(\frac{\sqrt{1-\varepsilon^2}-1}{\varepsilon}\right)^{\vert k \vert},
\label{coeffs}
\end{align}
for $G_0>0$, $\varepsilon^2<1$, and $k\in\mathbb{Z}$.
This result, by using Eq. (\ref{deff}), allows the determination of the explicit expression for the effective diffusion constant for this sinusoidal case, within the Fisk-Stratonovich interpretation
\begin{align}
    \overline{D}=G_0^2(1-\varepsilon^2).
    \label{dpart}
\end{align}
We also want to determine the full form of the probability density in this particular case. 
By recalling the definition of $\mathcal{D}(x)$, we obtain
{
\begin{align}
    \mathcal{D}(x)=\frac{1}{G_0\sqrt{1-\varepsilon^2}}\left\lbrace x +\frac{L}{2\pi i} \sum_{\begin{subarray}{c}
			k=-\infty\\
			k\neq 0
		\end{subarray}}^{+\infty}\frac{\beta^{\vert k \vert}}{k}\left[\exp{\left(\frac{2\pi i k x}{L}\right)}-1\right]\right\rbrace,
\end{align}}
where we introduced $\beta=(\sqrt{1-\varepsilon^2}-1)/\varepsilon$ to simplify the notation. For the following calculations, it is useful to note that $\beta^2<1$ if and only if $\varepsilon^2<1$. 
Moreover it is evident that 
\begin{align}
     \sum_{\begin{subarray}{c}
			k=-\infty\\
			k\neq 0
		\end{subarray}}^{+\infty}\frac{\beta^{\vert k \vert}}{k}=0,
\end{align}
and therefore the expression of $\mathcal{D}(x)$ is simplified as follows
{
\begin{align}
    \mathcal{D}(x)=\frac{1}{G_0\sqrt{1-\varepsilon^2}}\left\lbrace x +\frac{L}{2\pi i} \sum_{\begin{subarray}{c}
			k=-\infty\\
			k\neq 0
		\end{subarray}}^{+\infty}\frac{\beta^{\vert k \vert}}{k}\exp{\left(\frac{2\pi i k x}{L}\right)}\right\rbrace.
		\label{ddx}
\end{align}
We prove in Appendix \ref{appeB} that the series appearing in the expression for $\mathcal{D}(x)$ can be calculated in closed form. This development leads to the following final formula}
\begin{align}
\nonumber
    \mathcal{D}(x)=&\frac{1}{G_0\sqrt{1-\varepsilon^2}}\left\lbrace x +\frac{L}{\pi} \left[\arctan\left(\sqrt{\frac{1-\varepsilon}{1+\varepsilon}}\tan\frac{\pi x}{L}\right)\right.\right.\\
    &\left.\left.-\arctan\left(\tan\frac{\pi x}{L}\right)\right]\right\rbrace.
		\label{ddx2}
\end{align}
The latter result can be directly used  in the expression of the probability density
\begin{align}
    \rho(x,t)=\frac{\exp\left[-\frac{\mathcal{D}^2(x)}{4t}\right]}{G_0\sqrt{4\pi t}\left(1+\varepsilon\cos\frac{2\pi x}{L}\right)},
    \label{density}
\end{align}
in order to study the space and time evolution of the stochastic system. 

An example of application of this result can be seen in Fig. \ref{dens}, where we represent the density $\rho(x,t)$ as a function of $x$, and parameterised by $t$, for four different cases (with $L=1$): 
(i) $G_0=1$ and $\varepsilon=0.3$; (ii) $G_0=1$ and $\varepsilon=0.7$; (iii) $G_0=5$ and $\varepsilon=0.3$; (iv) $G_0=5$ and $\varepsilon=0.7$. 
First of all, we can observe that the general trend corresponds to the familiar one of classical diffusion (think of the Wiener process given by Eq. (\ref{propar}) with $g(x)$ constant) in which, as time increases, the density broadens on the spatial axis, increasing the variance.
This is true for any choice of parameters $G_0$ and $\varepsilon$.
However, in our case, we observe a perturbation to the classical Gaussian shape consisting of oscillations arising from heterogeneous periodic sinusoidal diffusion.
The points at which the density has local minima correspond to the maxima of the local diffusion function $g(x)$; reciprocally, the points at which the density has local maxima correspond to the minima of $g(x)$.
This is explained by the fact that the function $g(x)$ somehow represents the mobility of the particle that is moving, under overdamped assumptions, in the periodic diffusive system. 
In other words, it can be said that periodic diffusion induces particle trapping phenomena (corresponding to density maxima) in areas of low mobility, where the particle moves with more difficulty. 
Since in reality the local diffusion coefficient depends on both the temperature and the friction coefficient, these trapping phenomena can be achieved both with heterogeneous temperatures and with variable friction. 
This is in accord with the work of Landauer \cite{landauer1988}, discussed in the Introduction.
From Fig. \ref{dens}, we deduce that the evolution of the diffusion phenomenon is faster for increasing values of $G_0$ and slower for increasing values of $\varepsilon$.
This is consistent with the expression of the effective diffusion constant given in Eq. (\ref{dpart}): $D_{\it eff}$ is indeed increasing with $G_0$ and decreasing with $\varepsilon$.
We finally observe that the trapping phenomenon is particularly amplified for large values of $G_0$ and values of $\varepsilon$ close to 1.

\section{Periodic heterogeneous diffusion with arbitrary stochastic interpretation}
\label{general}

We now study the effective diffusion coefficient $D_{\it eff}$ for the stochastic differential equation with periodic heterogeneous diffusion in the case of a general discretization parameter $ 0 \leq \alpha \leq 1$.
{This means that we need to study the properties of Eq. (\ref{fpgenintro}). In particular,}
the standard notation for the propagator that solves Eq. (\ref{fpgenintro}) is given by $W=W(x,t;x_0,t_0)$. We use
the initial condition $W(x,t_0;x_0,t_0)=\delta(x-x_0)$ for defining this propagator. 
In the following calculations, we set the initial time $t_0=0$ and, in order to simplify the notation, use only the first two arguments for $W$ to denote it as $W(x,t)$, associated with the initial condition $W(x,0)=\delta(x-x_0)$. To further simplify the following development, we also introduce the quantities
\begin{align}
    \mathcal{A}(x)& \equiv \frac{1}{g^{2\alpha}(x)},\\
\mathcal{B}(x)& \equiv \frac{1}{g^{2(1-\alpha)}(x)},
\end{align}
that are both periodic functions with period $L$.
We therefore analyse the differential problem
\begin{align}
  \label{fptosolve}
      \frac{\partial W(x,t)}{\partial t}&=\frac{\partial}{\partial x}\left\lbrace\frac{1}{\mathcal{A}(x)}\frac{\partial }{\partial x}\left[\frac{1}{\mathcal{B}(x)}W(x,t)\right]\right\rbrace,\\
          W(x,0)&=\delta(x-x_0),
      \label{inittosolve}
\end{align}
with the aim of finding the effective diffusion constant $ D_{\it eff} $ in terms of $\mathcal{A}(x)$ and $ \mathcal{B}(x)$. If $\mathcal{A}$ and $\mathcal{B}$ were constant in space, then the diffusion coefficient would be $1/(\mathcal{A}\mathcal{B})$.
 
{For this general problem, it is probably impossible to obtain a closed form expression for the probability density $W(x,t)$, and for this reason 
we now use the definition of the effective diffusion constant given by}
\begin{align}
    D_{\it eff}  \equiv \frac{a}{2\overline{T}_{FP}},
    \label{second}
\end{align}
and will give a precise meaning of $a$ and 
$\overline {T}_{FP}$ for our problem. 

In this case, in addition to the initial condition $W(x,0)=\delta(x-x_0)$, the Fokker-Planck equation is considered to be equipped  with two adsorbing conditions at the points $x_0-a$ and $x_0+a$. Hence, the interval between the boundary conditions is symmetric with respect to the position of the initial condition; thus we have the supplementary conditions $W(x_0-a,t)=0$ and $W(x_0+a,t)=0$. The particle is adsorbed at these boundary points, and the density gradually tends to zero in the inner range as time passes. 
The quantity $\overline{T}_{FP}$ is the mean first passage time, i.e., the average value of the time it takes for the particle starting at $x_0$ to arrive at one of the two boundary points $x_0-a$ or $x_0+a$.  
In other words, it is the average time to wait for the particle to leave the considered interval or,
equivalently, the average time the particle needs to travel the length $a$ toward the left or toward the right, from the initial position $x_0$.

Since we are studying a periodic system, we choose the length $a$ equal to a multiple $n$ of the period $L$. 
Then 
\begin{align}
\label{leftbc}
    W(x_0-nL,t)&=0,\\
    W(x_0+nL,t)&=0\, .
    \label{rightbc}
\end{align}
Next we define the survival probability
\begin{align}
    S(t)=\int_{x_0-nL}^{x_0+nL}W(x,t)\mathrm{d} x,
\end{align}
which is the probability for the particle to be in the 
interval $(x_0-nL,x_0+nL)$.
The survival probability also represents the probability that the first passage time $T_{FP}$ exceeds a given time value $t$ \cite{gardiner2009,coffey2004}
\begin{align}
    \Pr\left\lbrace T_{FP}>t \right\rbrace= S(t)=\int_{x_0-nL}^{x_0+nL}W(x,t)\mathrm{d} x.
    \label{prst}
\end{align}
Note the first passage time $T_{FP}$, defined as the time to reach either extremity placed  at $x_0\pm a=x_0\pm nL$, is a random variable whose distribution law and density can be defined, as we will see shortly below. Through these, we can then define the mean first passage time, which is the key quantity in this procedure. 
Indeed, from Eq. (\ref{prst}) we have that
\begin{align}
    \Pr\left\lbrace T_{FP}\le t \right\rbrace=1-\Pr\left\lbrace T_{FP}>t \right\rbrace=1-\int_{x_0-nL}^{x_0+nL}W(x,t)\mathrm{d} x
\end{align}
is the probability distribution of the first passage time. The corresponding probability density is therefore given by
\begin{align}
    f(t) = \frac{\mathrm{d}}{\mathrm{d} t} \Pr\left\lbrace T_{FP}\le t \right\rbrace=-\int_{x_0-nL}^{x_0+nL}\frac{\partial W(x,t)}{\partial t} \mathrm{d} x.
\end{align}
We can calculate the mean first passage time as
\begin{align}
\nonumber
    \overline{T}_{FP}&=\int_0^{+\infty}tf(t)\mathrm{d} t=-\int_0^{+\infty}t\int_{x_0-nL}^{x_0+nL}\frac{\partial W(x,t)}{\partial t} \mathrm{d} x\mathrm{d} t\\
    &=-\int_{x_0-nL}^{x_0+nL}\int_0^{+\infty}t\frac{\partial W(x,t)}{\partial t} \mathrm{d} t \mathrm{d} x,
\end{align}
and an integration by parts on the internal time integral yields
\begin{align}
    \overline{T}_{FP}=\int_{x_0-nL}^{x_0+nL}\int_0^{+\infty}W(x,t) \mathrm{d} t \mathrm{d} x.
\end{align}
By introducing 
\begin{align}
    k(x)=\int_0^{+\infty}W(x,t) \mathrm{d} t ,
\end{align}
we obtain the final expression
\begin{align}
    \overline{T}_{FP}=\int_{x_0-nL}^{x_0+nL}k(x) \mathrm{d} x.
    \label{mfpt}
\end{align}
We now determine the function $k(x)$ for our problem stated in Eqs. (\ref{fptosolve}), (\ref{inittosolve}), (\ref{leftbc}) and (\ref{rightbc}).
We start by integrating Eq. (\ref{fptosolve}) over the time from 0 to $+\infty$, and we find
{
\begin{align}
    W(x,+\infty)-W(x,0)=\frac{\mathrm{d}}{\mathrm{d} x}\left\lbrace\frac{1}{\mathcal{A}(x)}\frac{\mathrm{d} }{\mathrm{d} x}\left[\frac{k(x)}{\mathcal{B}(x)}\right]\right\rbrace.
\end{align}
Here and in the sequel we use the differential operator $\frac{\mathrm{d}}{\mathrm{d} x}$ instead of $\frac{\partial}{\partial x}$ since the time variable is no longer present.} 
Since $W(x,+\infty)\to 0$ and $W(x,0)=\delta(x-x_0)$, as defined in Eq. (\ref{inittosolve}), we have an equation for $k(x)$
\begin{align}
    -\delta(x-x_0)=\frac{\mathrm{d}}{\mathrm{d} x}\left\lbrace\frac{1}{\mathcal{A}(x)}\frac{\mathrm{d} }{\mathrm{d} x}\left[\frac{k(x)}{\mathcal{B}(x)}\right]\right\rbrace,
    \label{eqfork}
\end{align}
for which the boundary conditions in Eqs. (\ref{leftbc}) and (\ref{rightbc}) can be rewritten as $k(x_0-nL)=0$ and $k(x_0+nL)=0$. 
{The solution of this equation is relegated to the Appendix \ref{appeC}, and the final result can be written as}
\begin{align}
     k(x)=\left\lbrace\begin{array}{cc}
          \frac{1}{2}\mathcal{B}(x)\int_{x_0-nL}^x\mathcal{A}(\xi)\mathrm{d} \xi, &  x_0-nL \le x < x_0, \\
          \frac{1}{2}\mathcal{B}(x)\int_{x}^{x_0+nL}\mathcal{A}(\xi)\mathrm{d} \xi, &  x_0 < x \le x_0+nL.
     \end{array}\right.
     \label{ksol}
\end{align}

Now, the mean first passage time can be explicitly calculated through Eq. (\ref{mfpt}), eventually yielding
\begin{align}
\nonumber
    \overline{T}_{FP}=&\frac{1}{2}\int_{x_0-nL}^{x_0}\mathcal{B}(x)\int_{x_0-nL}^x\mathcal{A}(\xi)\mathrm{d} \xi\mathrm{d} x\\
    &+\frac{1}{2}\int_{x_0}^{x_0+nL}\mathcal{B}(x)\int_{x}^{x_0+nL}\mathcal{A}(\xi)\mathrm{d} \xi\mathrm{d} x,
\end{align}
where, remember, $\mathcal{A}$ and $\mathcal{B}$ are periodic functions with period $L$.
{To simplify the obtained expression for $\overline{T}_{FP}$, we apply the substitution $y=x+nL$ to the first integral, and we get
\begin{align}
\nonumber
    \overline{T}_{FP}=&\frac{1}{2}\int_{x_0}^{x_0+nL}\mathcal{B}(y-nL)\int_{x_0-nL}^{y-nL}\mathcal{A}(\xi)\mathrm{d} \xi\mathrm{d} y\\
      &+\frac{1}{2}\int_{x_0}^{x_0+nL}\mathcal{B}(x)\int_{x}^{x_0+nL}\mathcal{A}(\xi)\mathrm{d} \xi\mathrm{d} x.
    \end{align}
    Now, the periodicity of $\mathcal{B}$ yields
    \begin{align}
        \overline{T}_{FP}=&\frac{1}{2}\int_{x_0}^{x_0+nL}\mathcal{B}(x)\left[\int_{x_0-nL}^{x-nL}\mathcal{A}(\xi)\mathrm{d} \xi+\int_{x}^{x_0+nL}\mathcal{A}(\xi)\mathrm{d} \xi\right]\mathrm{d} x,
     \end{align}
    and the periodicity of $\mathcal{A}$ leads to
    \begin{align}
        \overline{T}_{FP}=&\frac{1}{2}\int_{x_0}^{x_0+nL}\mathcal{B}(x)\left[\int_{x_0}^{x}\mathcal{A}(\xi)\mathrm{d} \xi+\int_{x}^{x_0+nL}\mathcal{A}(\xi)\mathrm{d} \xi\right]\mathrm{d} x.
    \end{align}
    Therefore, the mean first passage time is simplified as
     \begin{align}
    \nonumber
    \overline{T}_{FP}=&\frac{1}{2}\int_{x_0}^{x_0+nL}\mathcal{B}(x)\mathrm{d} x\int_{x_0}^{x_0+nL}\mathcal{A}(\xi)\mathrm{d} \xi\\
    =&\frac{n^2}{2}\int_{0}^{L}\mathcal{B}(x)\mathrm{d} x\int_{0}^{L}\mathcal{A}(\xi)\mathrm{d} \xi.
\end{align}
}
To conclude, we can use Eq. (\ref{second}) with $a=nL$, and we obtain the effective diffusion constant as
\begin{align}
D_{\it eff} = \frac{(nL)^2}{2\overline{T}_{FP}}=\frac{L^2}{\int_{0}^{L}\mathcal{B}(x)\mathrm{d} x\int_{0}^{L}\mathcal{A}(\xi)\mathrm{d} \xi}=\frac{1}{\left\langle \mathcal{A} \right\rangle\left\langle \mathcal{B} \right\rangle}.
\label{resres}
\end{align}
This is the key result of the present Section and, for the interested readers, we prove in the Appendix \ref{appeD} that it is fully consistent with a classical steady-state homogenization approach. 
Our result can be explicitly rewritten in terms of the function $g(x)$ from the definitions of $\mathcal{A}$ and $\mathcal{B}$
\begin{align}
D_{\it eff} = \frac{1}{\left\langle \frac{1}{g^{2\alpha}(x)} \right\rangle\left\langle \frac{1}{g^{2(1-\alpha)}(x)}\right\rangle},
\label{risddd}
\end{align}
where the symbol $\left\langle \cdot\right\rangle$ represents the mean value of the argument over a period. 
This result is consistent with the effective diffusion constant found in Section \ref{secstrato} for Wereide's law ($\alpha=1/2$), and for the well-known effective diffusion constant for Fick's law ($\alpha=1$). It also provides a new result for Chapman's diffusion law ($\alpha=0$), which coincides with the result concerning the Fick's law. Indeed, $D_{\it eff}$ in Eq. (\ref{risddd}) is invariant under the substitution $\alpha \rightleftarrows 1-\alpha$.

\begin{figure*}[ht!]
\includegraphics[width=8.5cm]{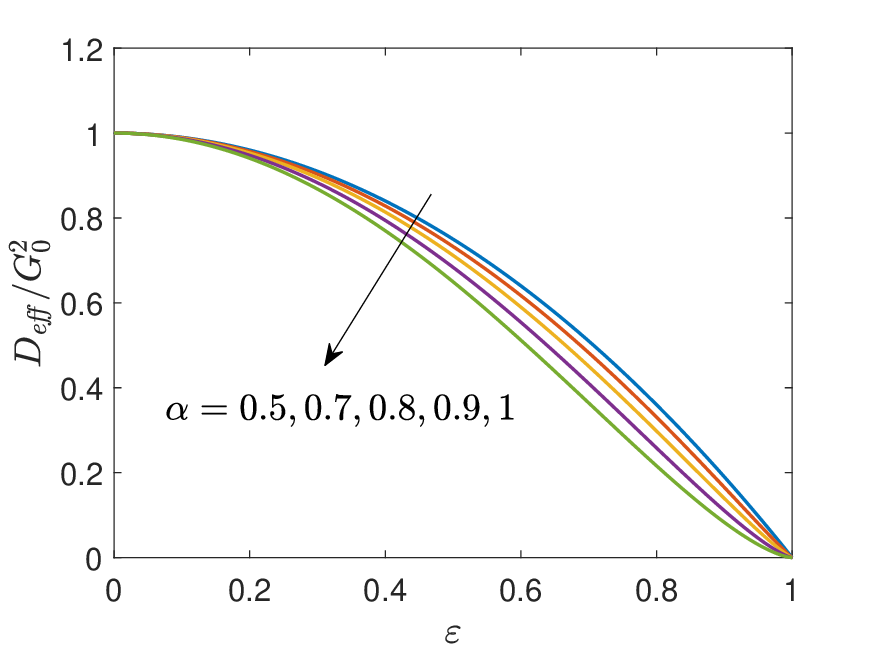}
\includegraphics[width=8.5cm]{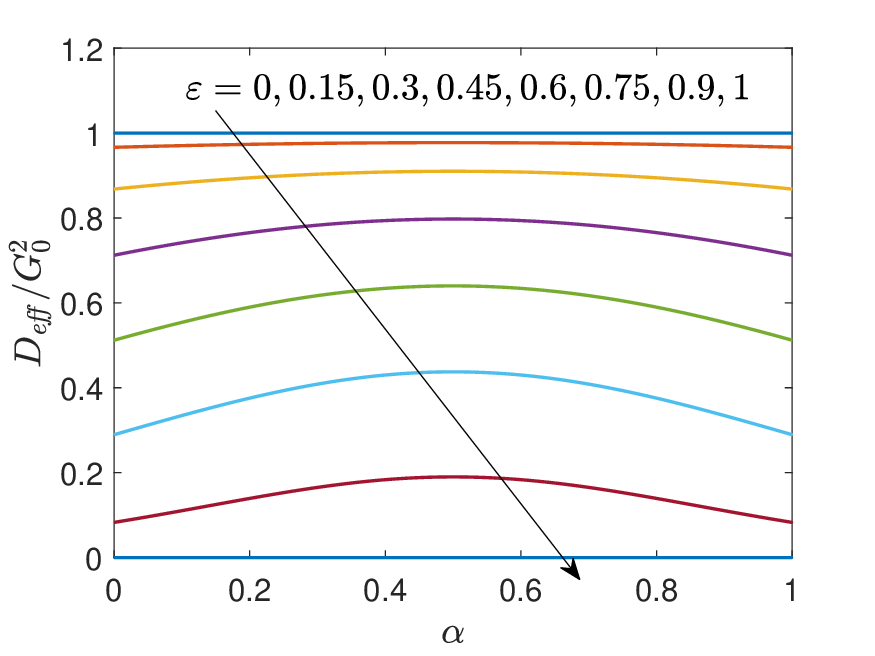}
\caption{\label{dtrig} Behavior of the effective diffusion constant with a periodic multiplicative noise described by $g(x)$ in Eq. (\ref{trig}). The behavior of $D_{\it eff}/G_0^2$ given in  Eq. (\ref{risdddtrig}) is represented versus $\varepsilon$ in the left panel and versus $\alpha$ in the right panel. }
\end{figure*}

We can now obtain the explicit result for the exemplary sinusoidal function $g(x)$ given in Eq. (\ref{trig}). By considering $\nu=2\alpha$ or $\nu=2(1-\alpha)$, we can calculate each of the average values in Eq. (\ref{risddd}) as
\begin{align}
    \left\langle \frac{1}{g^{\nu}(x)} \right\rangle=\frac{1}{L}\int_0^LG_0^{-\nu}\left(1+\varepsilon\cos\frac{2\pi x}{L}\right)^{-\nu}\mathrm{d} x,
\end{align}
where, as before, $G_0>0$ and $\varepsilon^2<1$. By the change of variable $\vartheta=2\pi x/L$, we obtain
{
\begin{align}
    \left\langle \frac{1}{g^{\nu}(x)} \right\rangle=\frac{1}{2\pi}\int_0^{2\pi} G_0^{-\nu}\left(1+\varepsilon\cos\vartheta\right)^{-\nu}\mathrm{d} \vartheta.
\end{align}
}
Now, we can introduce the parameters $k$ and $z$ defined in such a way that $kz=G_0$ and $k\sqrt{z^2-1}=G_0\varepsilon$. By straightforward calculations we have $k=G_0\sqrt{1-\varepsilon^2}$ and $z=1/\sqrt{1-\varepsilon^2}$, and therefore the average value assumes the form
{
\begin{align}
    \left\langle \frac{1}{g^{\nu}(x)} \right\rangle=\frac{1}{2\pi}\int_0^{2\pi} k^{-\nu}\left(z+\sqrt{z^2-1}\cos\vartheta\right)^{-\nu}\mathrm{d} \vartheta.
\end{align}
}
We can remember the Legendre function representation by means of the following Laplace integral \cite{gradshteyn1965,abramowitz1970}
\begin{align}
    P_\mu(z)=\frac{1}{2\pi}\int_0^{2\pi}\left(z+\sqrt{z^2-1}\cos\vartheta\right)^\mu\mathrm{d} \vartheta,
\end{align}
which can be used with both the argument $z$ and the  order $\mu$ in the real domain, and we get
\begin{align}
    \left\langle \frac{1}{g^{\nu}(x)} \right\rangle= k^{-\nu}P_{-\nu}(z)=\frac{P_{-\nu}\left(\frac{1}{\sqrt{1-\varepsilon^2}}\right)}{\left(G_0\sqrt{1-\varepsilon^2}\right)^{\nu}}.
\end{align}
Finally, the expression for the effective diffusion constant in this sinusoidal case is given by the closed form expression 
\begin{align}
D_{\it eff} =\frac{G_0^2(1-\varepsilon^2)}{P_{-2\alpha}\left(\frac{1}{\sqrt{1-\varepsilon^2}}\right)P_{-2(1-\alpha)}\left(\frac{1}{\sqrt{1-\varepsilon^2}}\right)},
\label{risdddtrig}
\end{align}
which shows explicitly the dependence of $D_{\it eff}$ on the discretization parameter $\alpha \in [0,1]$ and the perturbation amplitude $\varepsilon$ (with $\varepsilon^2<1$).
Of course, $D_{\it eff}$ is an even function of $\varepsilon$. 
We observe that $D_{\it eff} = G_0^2$ when $\varepsilon=0$, as expected, since $P_\mu(1)=1\, \forall\, \mu$.
Moreover, considering that $P_{-1}(z)=P_0(z)=1$, and $P_{-2}(z)=P_1(z)=z$, we have the following particular results: (i) for $\alpha=1/2$, we get $D_{\it eff} = G_0^2(1-\varepsilon^2)$, in agreement with Eqs. (\ref{deff}), (\ref{coeffs}), and (\ref{dpart}); (ii) for $\alpha=0$ and $\alpha=1$, we get the identical result $D_{\it eff} = G_0^2(1-\varepsilon^2)^{3/2}$.

To further show the relationship between $D_{\it eff}$ and $\alpha$, we can determine the approximation of the obtained result for small values of $\varepsilon$. It means that we consider small periodic perturbation of the function $g(x)$, see Eq. (\ref{trig}).
To this aim, we evaluate the Legendre function for small values of $\varepsilon$, as follows
\begin{align}
\nonumber
    P_\mu(z)=&\frac{z^\mu}{2\pi}\int_0^{2\pi}\left(1+\frac{\sqrt{z^2-1}}{z}\cos\vartheta\right)^\mu\mathrm{d} \vartheta\\
    \nonumber
    =&\frac{z^\mu}{2\pi}\int_0^{2\pi}\left(1+\frac{\sqrt{z^2-1}}{z}\mu\cos\vartheta\right.\\
     &\left.+\frac{1}{2}\mu(\mu-1)\frac{{z^2-1}}{z^2}\cos^2\vartheta+...\right)\mathrm{d} \vartheta,
\end{align}
where we applied the Newton binomial theorem. Performing the integration and recalling that $z=1/\sqrt{1-\varepsilon^2}$, we get the second-order expansion
\begin{align}
    P_\mu\left(\frac{1}{\sqrt{1-\varepsilon^2}}\right)=1+\frac{1}{4}\mu(\mu+1)\varepsilon^2+...,
\end{align}
which is valid for small values of $\varepsilon$.
We can use this approximation in Eq. (\ref{risdddtrig}), and we obtain
\begin{align}
D_{\it eff} =G_0^2\left[1-\frac{1}{2}\left(3-4\alpha+4\alpha^2\right)\varepsilon^2+...\right],
\label{risdddtrigappr}
\end{align}
which shows even more explicitly the dependence of the diffusion constant on the parameter $\alpha$, also for small perturbations of $g(x)$. Importantly, we observe that $D_{\it eff} \le G_0^2$ since $3-4\alpha+4\alpha^2$ is positive for any value of $\alpha$. From a physical point of view, this means that periodically perturbing a diffusion coefficient means decreasing its effective value. We also remark that the polynomial $3-4\alpha+4\alpha^2$ is invariant to the substitution $\alpha \rightleftarrows 1-\alpha$, as expected, since the approximated relation given in Eq. (\ref{risdddtrigappr}) has been derived from the general solution stated in Eq. (\ref{risddd}).

A numerical implementation of Eq. (\ref{risdddtrig}) is presented in Fig. \ref{dtrig}, where we represent $D_{\it eff}/G_0^2$ versus $\varepsilon$ and $\alpha$.
In the left panel we find the effective diffusion constant as a function of $\varepsilon$ and parameterized by the discretization coefficient $\alpha$. 
Reciprocally, in the right panel we find the effective diffusion constant as a function of $\alpha$ and parameterized by the amplitude $\varepsilon$ of the periodic perturbation.
The ranges of variation of these parameters are clearly specified within the panels.
In the left panel we see that the effective diffusion coefficient decreases monotonically with $\varepsilon$ and reaches zero when $\varepsilon=1$.
This decrease is explained by the fact that as $\varepsilon$ increases, areas with weak local diffusion appear and thus the particle moves less easily, thus reducing the effective diffusion constant. 
In the limit of $\varepsilon=1$, the local diffusion becomes zero at certain points that the particle will be prevented from passing through. 
From a physical point of view, the first-order stochastic differential equation can be thought of as corresponding to overdamped motion, and so the absence of inertial effects helps even more to understand this cancellation of diffusion.
Regarding the effects of the discretization parameter $\alpha$, we observe that when it varies between 1/2 and 1, the effective diffusion constant is monotonically decreasing. 
It is easily seen that this result is consistent with the approximate formula given in Eq. (\ref{risdddtrigappr}).
This point will be further commented in the next Section.
In the right panel we can come to the same conclusions, and we also see the graphical representation of the symmetry induced by the invariance of the result under the substitution $\alpha \rightleftarrows 1-\alpha$.
This finally allows us to observe that for any fixed value of $\varepsilon$, the maximum value of effective diffusion constant is obtained for $\alpha=1/2$, that is, in the stochastic Fisk-Stratonovich interpretation (or Wereide's diffusion law).

\section{Generalized Lifson-Jackson theorem}
\label{lija}

The result obtained in the previous Section is closely related to the Lifson-Jackson theorem concerning the effective diffusion constant for a particle embedded in a periodic potential energy \cite{lifson1962}.
We briefly summarize the result of this theorem in order to develop a generalization to the case of heterogeneous diffusion superimposed on the effects of the potential energy. 

For a particle that is experiencing an overdamped motion under the effect of any potential energy $U(x)$, we can write the first order Langevin's equation
\begin{align}
    \frac{\mathrm{d} x}{\mathrm{d} t}=-\frac{1}{m\gamma}\frac{\mathrm{d} U}{\mathrm{d} x}+\sqrt{\frac{k_B T}{m\gamma}}\xi(t),
\end{align}
where $m$ is the particle mass, $\gamma$ is its friction coefficient, $T$ is the temperature, and $k_B$ is  Boltzmann's constant. 
For now, the value of $\sqrt{\frac{k_B T}{m\gamma}}$ is constant, and then we have a standard additive noise. 
The associated Fokker-Planck (or Smoluchovski) equation can be written as
\begin{align}
    \frac{\partial W}{\partial t}=\frac{1}{m\gamma}\frac{\partial }{\partial x}\left(\frac{\partial U}{\partial x}W\right)+\frac{k_B T}{m\gamma}\frac{\partial^2 W }{\partial x^2},
    \label{smolu}
\end{align}
which is equivalent to the more useful form
\begin{align}
    \frac{\partial W}{\partial t}=\frac{k_B T}{m\gamma}\frac{\partial }{\partial x}\left[e^{-\frac{U}{k_B T}}\frac{\partial }{\partial x}\left(e^{+\frac{U}{k_B T}}W\right)\right].
    \label{fppot}
\end{align}
Incidentally, this second form is well adapted to recognize that the asymptotic solution is correctly given by the Gibbs-Boltzmann distribution $W(x,+\infty)=e^{-\frac{U}{k_B T}}/Z$, where the partition function $Z$ it is used to normalize the probability density.
For this problem, with periodic potential energy and homogeneous diffusion constant, the effective diffusion constant is given by 
\begin{align}
    D_{\it eff} = \frac{\frac{k_B T}{m\gamma}}{\left\langle e^{-\frac{U(x)}{k_B T}} \right\rangle \left\langle e^{+\frac{U(x)}{k_B T}} \right\rangle},
    \label{lj}
\end{align}
where $k_B T/{m\gamma} $ is the diffusion constant of the homogeneous system with $U=0$, see Eq. (\ref{smolu}).

In their original paper \cite{lifson1962}, Lifson and Jackson proved their theorem through a method proposed by Pontryagin, Andronow and Witt that provides a differential equation for the average time taken by a particle moving under the combined effect of thermal agitation and a stationary force field to reach a given boundary \cite{pontryagin1933}.
Given the elegance and application importance of this theorem, many different demonstrations have been presented in the literature. 
For instance, the concept of mean first passage time has been used in Refs.\cite{weaver1979,weaver1984,bolanos2015}, a thorough technique based on an eigenfunction expansion applied to the Smoluchovski equation has been proposed in Ref.\cite{festa1978}, and a clever demonstration based on a few physical arguments has been developed in Ref.\cite{gunther1979}. In all these developments, the noise is always additive and therefore there is no influence of the discretization parameter $\alpha$.

Here, we propose a generalization taking into account a heterogeneous diffusion generated by a spatially varying friction coefficient $\gamma(x)$, and an arbitrary discretization coefficient $\alpha$.
In this situation, the Langevin equation assumes the form
\begin{equation} 
    \frac{\mathrm{d} x}{\mathrm{d} t}=h(x)+g(x)\xi(t),
    \label{langen}
\end{equation}
where  $h(x)$ and $g(x)$ are defined as follows
\begin{align}
\label{hdx}
  h(x)&=-\frac{1}{m\gamma(x)}\frac{\mathrm{d} U(x)}{\mathrm{d} x},\\
  g(x)&=\sqrt{\frac{k_B T}{m\gamma(x)}},
  \label{gdx}
\end{align}
and where $U(x)$ and $\gamma(x)$ are periodic with period $L$. 
It is easily seen that the heterogeneous diffusion is induced by the variable periodic friction coefficient.

The probability density $W(x,t)$ is the solution of the  Fokker-Planck equation \cite{risken1989,oksendal2003,gardiner2009,coffey2004,denisov2003,denisov2009,denisov2014}
\begin{equation}
    \frac{\partial W}{\partial t}=-\frac{\partial}{\partial x}\left[\left(h+2\alpha g\frac{\partial g}{\partial x}\right)W\right]+\frac{\partial^2}{\partial x^2}\left(g^2W\right),
\end{equation}
which can be rewritten as
\begin{align}
    \frac{\partial W}{\partial t}=\frac{\partial }{\partial x}\left[g^{2\alpha}e^{-\frac{U}{k_B T}}\frac{\partial }{\partial x}\left(g^{2(1-\alpha)}e^{+\frac{U}{k_B T}}W\right)\right].
\end{align}
This is a crucial result for the continuation of the discussion.
The equivalence between these two forms can be proved by performing the derivatives and by recalling the definitions of $h(x)$ and $g(x)$ in Eqs.(\ref{hdx}) and (\ref{gdx}).
We remark that, if $g$ is constant (i.e., $\gamma$ is constant), we retrieve Eq. (\ref{fppot}), and if $U=0$, we retrieve Eq. (\ref{fpgenintro}).
Now, this new form of the Fokker-Planck equation can be compared with Eq. (\ref{fptosolve}) provided that we introduce the new variable coefficients
\begin{align}
    \mathcal{A}(x)&=\frac{1}{g^{2\alpha}(x)e^{-\frac{U(x)}{k_B T}}},\\
\mathcal{B}(x)&=\frac{1}{g^{2(1-\alpha)}(x)e^{+\frac{U(x)}{k_B T}}},
\end{align}
and therefore we can directly write down the expression for the effective diffusion constant
\begin{align}
\nonumber
    D_{\it eff} = &\frac{1}{\left\langle \frac{e^{-\frac{U(x)}{k_B T}}}{g^{2(1-\alpha)}(x)} \right\rangle\left\langle \frac{e^{+\frac{U(x)}{k_B T}}}{g^{2\alpha}(x)} \right\rangle}\\
    =&\frac{1}{\left\langle \left[\frac{m\gamma(x)}{k_B T}\right]^{1-\alpha}e^{-\frac{U(x)}{k_B T}} \right\rangle\left\langle \left[\frac{m\gamma(x)}{k_B T}\right]^{\alpha}e^{+\frac{U(x)}{k_B T}} \right\rangle},
    \label{findpot}
\end{align}
which generalizes the Lifson-Jackson theorem stated in Eq. (\ref{lj}).
This result takes into consideration the combined effects of the periodic potential energy and the periodic heterogeneous diffusion, introduced through the variable friction coefficient.
Of course, this result still depends on  $\alpha$, but we have lost the symmetry induced by the substitution $\alpha \rightleftarrows 1-\alpha$. 
\begin{figure*}[t!]
\includegraphics[width=8.5cm]{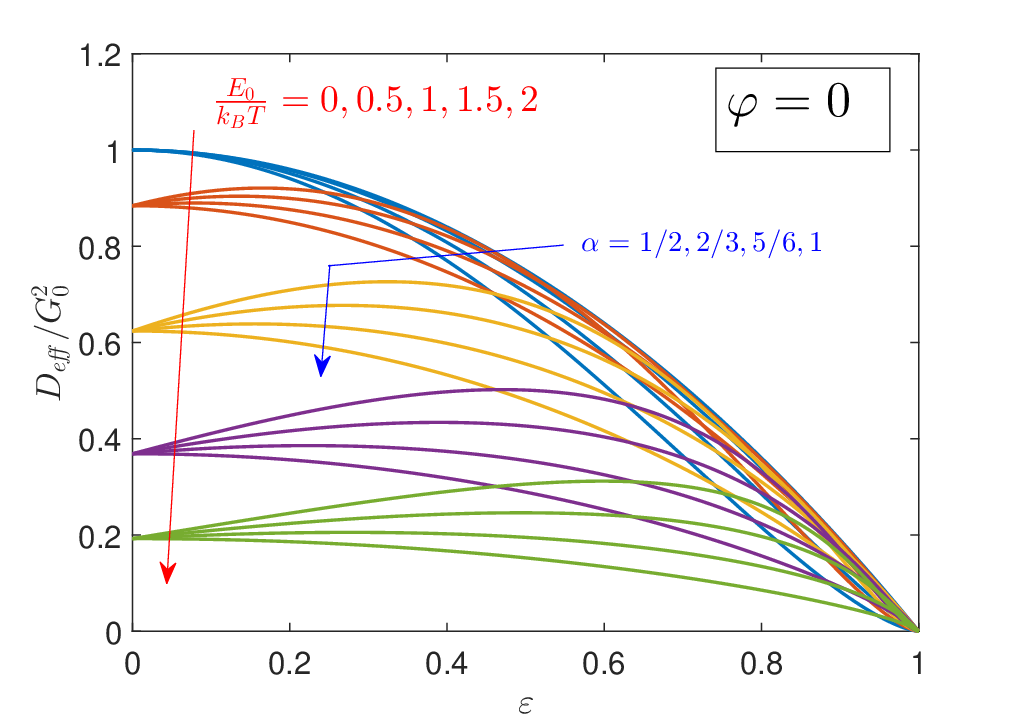}
\includegraphics[width=8.5cm]{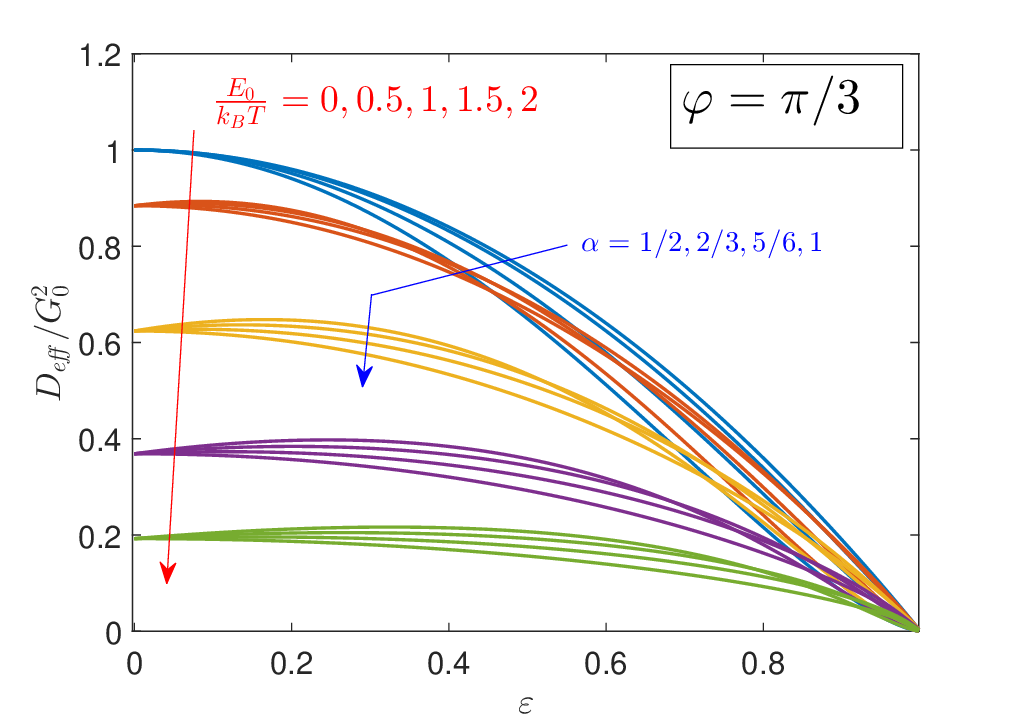}
\includegraphics[width=8.5cm]{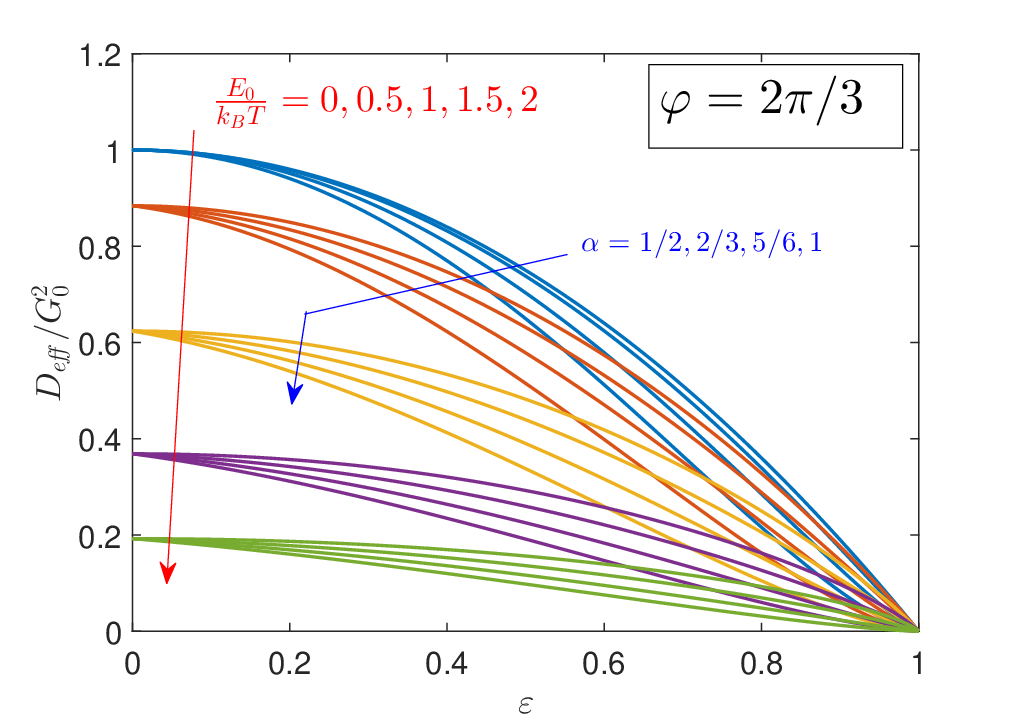}
\includegraphics[width=8.5cm]{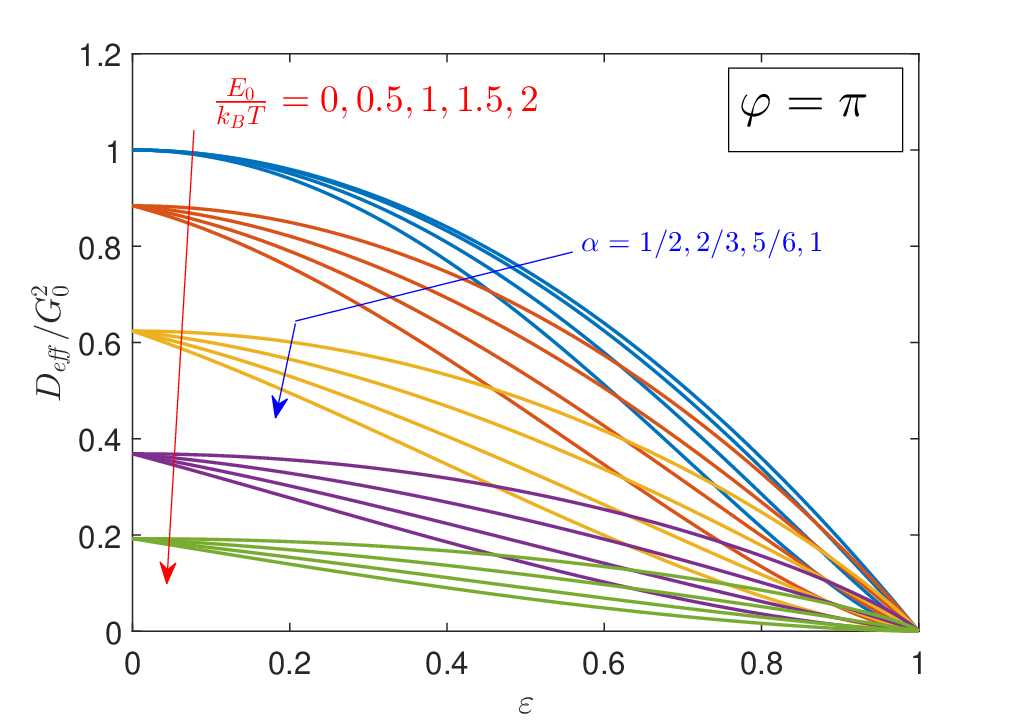}
\caption{\label{dtrigpot} Behavior of the effective diffusion constant with a periodic multiplicative noise described by $g(x)$ in Eq. (\ref{trig1}) and by a periodic potential described by $U(x)$ in Eq. (\ref{trig2}). The behavior of $D_{\it eff}/G_0^2$ given in  Eq. (\ref{findpot}) is represented versus $\varepsilon$ and parameterized by the energy ratio $E_0/(k_B T)$ and the discretization coefficient $\alpha$.}
\end{figure*}

An interesting general property of this result can be derived by means of the Cauchy-Schwartz inequality 
\begin{align}
    \left[\int_0^Lf(x)\ell(x)\mathrm{d} x\right]^2\le\int_0^Lf^2(x)\mathrm{d} x\int_0^L\ell^2(x)\mathrm{d} x,
\end{align}
which can be directly applied to the denominator of Eq. (\ref{findpot}).
Straightforward calculations deliver \begin{align}
    D_{\it eff} \le\frac{1}{\left\langle\frac{1}{g} \right\rangle^2 },
    \label{schwartz}
\end{align}
which means that the diffusion coefficient, regardless of potential energy and stochastic interpretation, is always smaller or equal than that corresponding to Wereide's diffusion (Fisk-Stratonovich interpretation) without drift. 
This can be easily seen in the right panel of Fig. \ref{dtrig}, where each curve shows a maximum point for $\alpha=1/2$.
Moreover, if $g(x)$ is constant, i.e. $\gamma(x)$ is constant, we obtain that $D_{\it eff} \le k_B T/(m\gamma )$ for any potential energy $U(x)$. It means that any shape of the periodic potential energy can only reduce the effective diffusion constant of the system with respect to the case with $U=0$.

\begin{figure*}[t!]
\includegraphics[width=8.5cm]{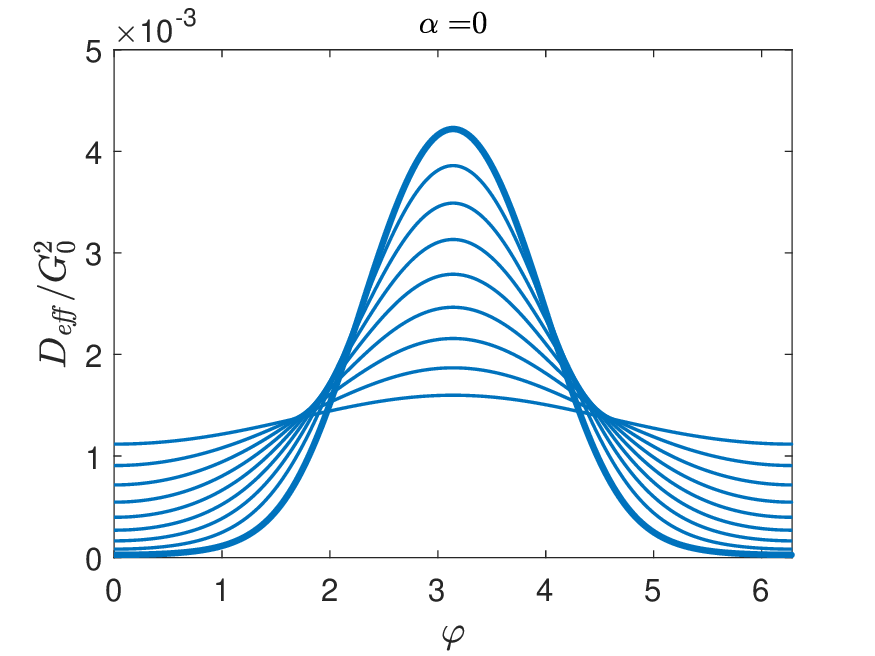}
\includegraphics[width=8.5cm]{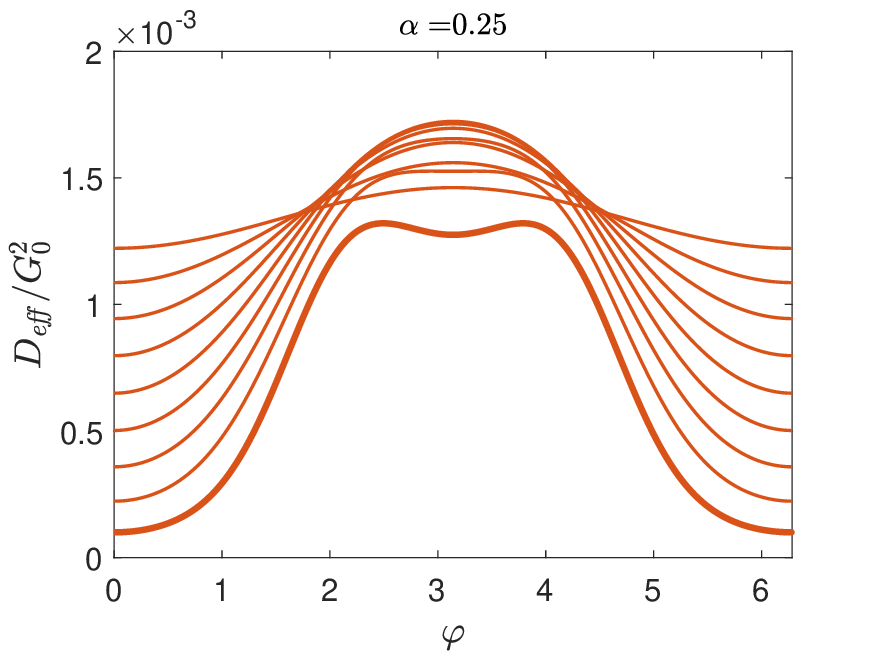}\\
\includegraphics[width=8.5cm]{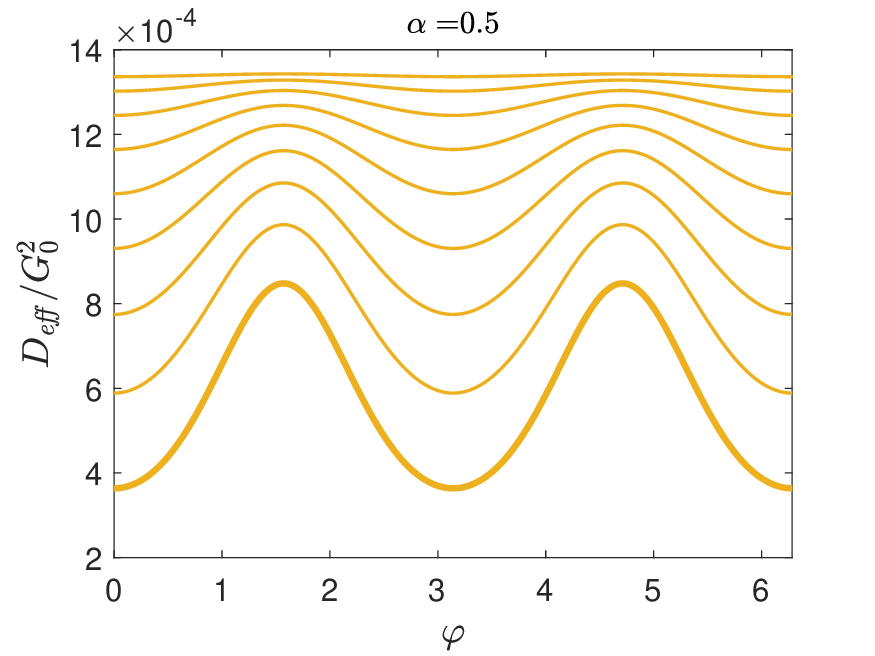}\\
\includegraphics[width=8.5cm]{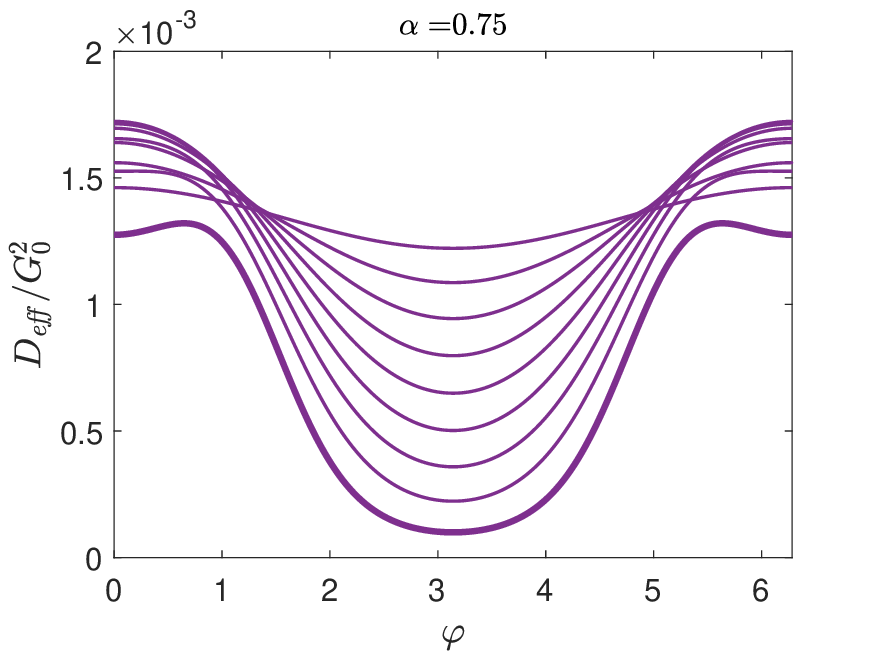}
\includegraphics[width=8.5cm]{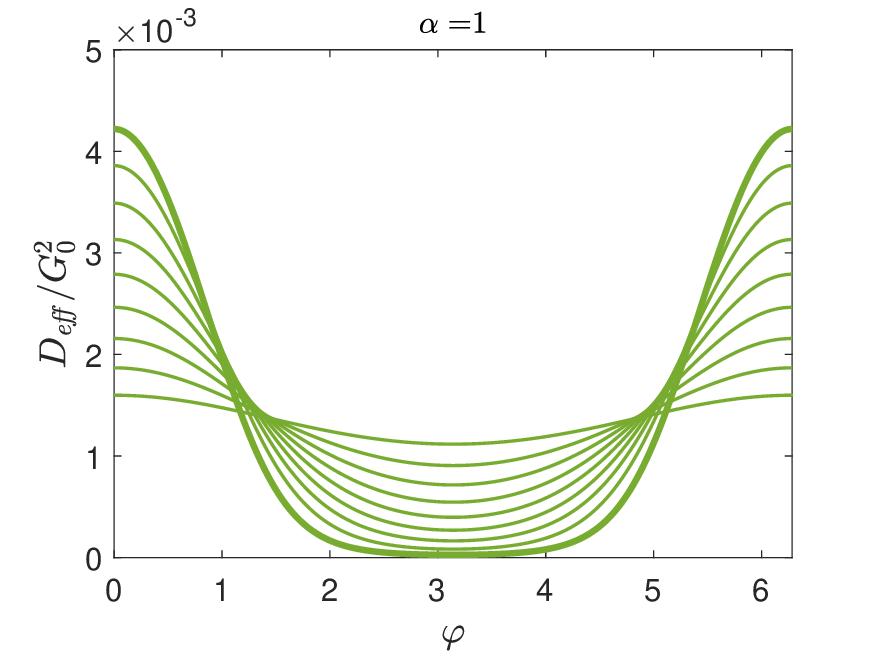}
\caption{\label{dtrigpotphase} Behavior of the effective diffusion constant with a periodic multiplicative noise described by $g(x)$ in Eq. (\ref{trig1}) and by a periodic potential described by $U(x)$ in Eq. (\ref{trig2}). The behavior of $D_{\it eff}/G_0^2$ given in  Eq. (\ref{findpot}) is represented versus $\varphi$ and parameterized by $\varepsilon=0.1,0.2,0.3,...,0.9$ (the thickest line corresponds to $\varepsilon=0.9$). We adopted $E_0/(k_B T)=5$ and the five different values of the discretization coefficient $\alpha$.}
\end{figure*}

We represent some numerical results concerning the effective diffusion constant given in Eq.(\ref{findpot}).
To do this, we choose the following profiles of heterogeneous diffusion and periodic potential
\begin{align}
    g(x)&=G_0\left(1+\varepsilon\cos\frac{2\pi x}{L}\right),
    \label{trig1}\\
    U(x)&=E_0\cos\left(\frac{2\pi x}{L}+\varphi\right),
    \label{trig2}
\end{align}
where $\varphi$ represents a phase shift between the two sinusoidal shapes. 
{ Experimental systems where the phase shift between drift and noise is achievable can be superlattices where the effect of friction (noise) is decoupled from the applied fields (drift). A particular case is represented by Moiré superlattices whose mechanical friction properties are well-known \cite{yan2024}.}
Obviously, the potential energy is defined except for an arbitrary additive constant that has no effect on the value of $D_{\it eff}$, as is easy to see from Eq. (\ref{findpot}).
In this case, it is not possible to calculate in closed form the effective constant $D_{\it eff}$, and the following results are therefore obtained with standard procedures for numerical integration.
In Fig. \ref{dtrigpot}, we show the behavior of $D_{\it eff}$ for four different values of the phase shift $\varphi=0, \pi/3, 2\pi/3, \pi$, in order to better explore the response of the system.
In each plot, we have reported five groups of curves, identified by different colours, corresponding to different values of the energy ratio ${E_0}/(k_B T)=0, 1/2,1, 3/2,2 $.
In each of these groups, the four curves of the same color correspond to different values of $\alpha=1/2, 2/3, 5/6, 1$. 
To identify these curves, we remark that the topmost curve in each group always corresponds to the  value $\alpha=1/2$, and the bottom curve always corresponds to the  value $\alpha=1$.
We can observe the following general trends that explain the dynamics of the system. First, in contrast to the case without drift, as $\varepsilon$ increases, we observe in some curves  an initial increase followed by a decrease in the effective diffusion constant. This is due to the interaction between potential energy and heterogeneous diffusion. 
When $\varepsilon=1$, we observe that in each case such an effective diffusion is zero because the particle cannot go through the points where the local diffusion coefficient is zero.  
As before, this point can be interpreted with an overdamped particle motion, where all inertial effects are completely neglected.
We also observe that the  energy ratio $E_0/(k_BT)$ strongly influences the evolution of the effective diffusion constant. Of course, when $E_0=0$ we get the same results as in Fig. \ref{dtrig} (left panel), which correspond to the blue groups of curves in Fig. \ref{dtrigpot}. 
In fact, in this case the drift is not present and only the heterogeneous diffusion acts on the system. 
Increasing the value of $E_0$ with respect to the thermal energy $k_BT$ shows a sharp decrease in effective diffusion justified by the fact that the particle in this case must cross an energy barrier of increasing amplitude. 
The values of $D_{\it eff}/G_0^2$ for $\varepsilon=0$ correspond the the original Lifson-Jackson theorem stated in Eq. (\ref{lj}), and the observed decreasing trend with the increasing ratio $E_0/(k_B T)$ corresponds to the previous analysis based on the Cauchy-Schwartz inequality.
The behavior induced by the discretization coefficient $\alpha$ is analogous to the one already observed in the case without drift, and can be summarized by saying that when alpha increases from 1/2 to 1 the coefficient $D_{\it eff}$ is monotonically decreasing, in agreement with the inequality given in Eq. (\ref{schwartz}). 
In addition, one is reminded of the symmetry of the response described by the invariance to  the interchange  $\alpha \rightleftarrows 1-\alpha$.
The analysis of the effects of the  phase shift $\varphi$ is more intricate because it represents the interaction between the profiles of heterogeneous diffusion and periodic potential. 
We can simply observe that the curvature of the response in the $(D_{\it eff}/G_0^2,\varepsilon)$ plane decreases as the phase shift increases from 0 to $\pi$. 
In fact, the maximum value of the effective diffusion constant gradually decreases as $\varphi$ increases between 0 and $\pi$.
This behavior is reproduced periodically for values of the phase shift outside this range.
A better explanation of the behavior of the diffusion constant versus the phase shift between drift and diffusion can be found in Fig. \ref{dtrigpotphase}.
Here, the behavior of $D_{\it eff}/G_0^2$, given in  Eq. (\ref{findpot}), is represented versus $\varphi$ and parameterized by $\varepsilon=0.1,0.2,0.3,...,0.9$ (the thickest line corresponds to $\varepsilon=0.9$). We adopted $E_0/(k_B T)=5$, and the five different values of the discretization coefficient $\alpha$.
The particular case with $\alpha=1/2$, corresponding to the Fisk-Stratonovich integration rule, can be interpreted as follows. 
To begin, we can state that
overcoming an energetic barrier is easier if the maximum of the noise occurs where the external force is largest. In this case, the noise would help to bring the particle over the energetic barrier. 
When $\alpha=1/2$, we see that the effective diffusion constant is larger if the phase shift is $\pi/2$ or $3\pi/2$.
It is easily seen that at these two points a maximum of $g(x)$ always corresponds to the point at which there is a maximum of force. This explains the behavior of the graphs for $\alpha=1/2$.
This interpretation is perfectly consistent with the findings of Landauer \cite{landauer1988}, and Breoni et al. \cite{breoni2022}.
It is then seen that the behavior is opposite in It\^o-type and anti-It\^o-type regions.
In fact, in the It\^o-type region we see a maximum of diffusivity for $\varphi=\pi$, while in the anti-It\^o-type region we see a minimum of diffusivity for $\varphi=\pi$. 
The observation of the five panels in Fig. \ref{dtrigpotphase} shows a continuity in behavior with respect to $\alpha$, described by the maximum of the It\^o trend and the minimum of the anti-It\^o trend, which are split into two maxima when $\alpha$ tends to 1/2.
However, the behavior away from Fisk-Stratonovich type region remains difficult to be interpreted physically because the stochastic integration does not follow the conventional rules of mathematical analysis.

\section{Conclusions}
\label{secconc}

In this work we have addressed the calculation of the
effective diffusion constant for stochastic processes
described by Fokker-Planck equations for different 
discretizations of the underlying Langevin equation.
The discretization parameter { $0 \leq \alpha \leq 1$}
enters the Fokker-Planck equation as a simple parameter.
We have determined the effective diffusion constant 
specifically for heterogeneous diffusions in the presence of a periodically modulated noise. 
First, we addressed such stochastic processes in the Fisk-Stratonovich interpretation by two different approaches. In this case with $\alpha=1/2$, the Fokker-Planck equation is typically referred to as Wereide's equation \cite{wereide1914}, and it is different from the more classical Fick's law \cite{fick1855,crank1975}. 
It is interesting to remark that the Fisk-Stratonovich interpretation \cite{fisk1963,stratonovich1966}, which is the most used at least in physical applications, leads to a diffusion equation not corresponding to the widely adopted Fick's law. 
Anyway, our approach provided the result 
$D_{\it eff}={\left\langle{1}/{g} \right\rangle^{-2}}$ for the effective diffusion constant.
If we now consider the H\"anggi-Klimontovich interpretation of the stochastic calculus with $\alpha=1$ 
\cite{haenggi1982,klimontovich1995}, in the associated Fokker-Planck equation
we immediately recognize the Fick's law \cite{fick1855}.
In this more classical case, the current probability density $J=-g^2\frac{\partial W}{\partial x}$ is given by the heterogeneous diffusion constant $g^2$ times the gradient of the density (with opposite sign), as typically occurs in the transport of heat, mass, or electrical charge \cite{crank1975,wang2008}. For this Fick's law, the effective diffusion constant is given by 
$D_{\it eff}=\left\langle{1}/{g^2} \right\rangle^{-1}$, as can be shown by different homogenization techniques applied to one-dimensional or stratified media \cite{milton2004,kim2011,camacho2013,giordano2014a,giordano2014b}. 
Therefore, the effective behavior of Fick's law is different from that of Wereide's law. 
The value of $\alpha$ is therefore of central importance in determining the effective diffusion constant. 
Moreover, to the authors' knowledge, no homogenization law was known for Chapman's diffusion, corresponding to $\alpha=0$ (Itô  stochastic interpretation \cite{ito1950}). 
For these reasons, we have discussed the case of general $\alpha$, which includes the three previous particular situations. 
The problem has been approached by determining the mean first passage time for a problem with adsorbing boundary conditions, which are symmetric with respect to the initial condition. 
In the Appendix D, we have also provided an equivalent demonstration based on standard homogenization techniques. 
We obtained an $\alpha$-dependent result for the diffusion constant, namely
$D_{\it eff} = \left\langle {1}/{g^{2\alpha}} \right\rangle^{-1}\left\langle {1}/{g^{2-2\alpha}}\right\rangle^{-1}$.
This expression is invariant under the substitution $\alpha \rightleftarrows 1-\alpha$, and therefore yields the same result for Fick's and Chapman's diffusion law. 
Moreover, Wereides's diffusion law always gives the highest value of the diffusion constant.
As an illustration of our results, we have applied them to a simple sinusoidal case, which can be treated in analytical detail. 
Interesting, the effective diffusion constant in this case can be written in terms of Legendre functions.
Finally, we have formulated a generalization of the Lifson-Jackson theorem for the case of combined periodic
potentials and diffusion coefficients. 
We discussed an example where potential energy and heterogeneous diffusion are both sinusoidal, with a variable phase shift. 
In this case, the calculation cannot be performed analytically and therefore we applied classical numerical techniques. 
The results have been discussed in terms of physical parameters describing the system.
Note in particular the importance of the discretization parameter $\alpha$ when studying the dependence of $D_{\it eff}$ on the shift angle $\varphi$.
An interesting perspective of our work concerns the generalization of the obtained results to the case of a { biased or tilted periodic potential energy (or washboard potential). }
It means that in addition to the periodic potential, one could consider a linear term representing a superimposed constant external force. In this case, we can define an average drift velocity and an effective diffusion constant. 
While this problem has been largely studied for homogeneous diffusion \cite{costantini1999,reimann2001,reimann2002,reimann2008,berezhkovskii2019,spiechowicz2020}, the case with heterogeneous diffusion still needs to be  investigated in more detail.

\appendix

{
\section{Alternative derivation of the diffusion constant in the Fisk-Stratonovich interpretation}

\label{appeA}

We propose here an alternative derivation of Eq. (\ref{deff}).} Since 
\begin{align}
\frac{\partial}{\partial x}\exp\left[-\frac{\mathcal{D}^2(x)}{4t}\right]=-\frac{\mathcal{D}(x)\mathcal{D}'(x)}{2t}   \exp\left[-\frac{\mathcal{D}^2(x)}{4t}\right], 
\end{align}
we can write the effective diffusion  constant defined in Eq. (\ref{dave1}) as
\begin{eqnarray}
    \overline{D}=\lim_{t\to \infty}\int_{-\infty}^{+\infty}\left[-\frac{x^2}{\mathcal{D}(x)}\frac{\frac{\partial}{\partial x}\exp\left[-\frac{\mathcal{D}^2(x)}{4t}\right]}{\sqrt{4\pi t}}\right]\mathrm{d} x.
    \label{dave3}
\end{eqnarray}
Now, integrating by parts we easily obtain
\begin{eqnarray}
\nonumber
    \overline{D}&=&\lim_{t\to \infty}\int_{-\infty}^{+\infty}\frac{\partial}{\partial x}\left(\frac{x^2}{\mathcal{D}(x)}\right)\frac{\exp\left[-\frac{\mathcal{D}^2(x)}{4t}\right]}{\sqrt{4\pi t}}\mathrm{d} x\\
    &=&\lim_{t\to \infty}\int_{-\infty}^{+\infty}\frac{2x\mathcal{D}(x)-x^2\mathcal{D}'(x)}{\mathcal{D}^2(x)}\frac{\exp\left[-\frac{\mathcal{D}^2(x)}{4t}\right]}{\sqrt{4\pi t}}\mathrm{d} x.\,\,\,\,\,\,\,\,
    \label{dave4}
\end{eqnarray}
First of all, we have to prove that there are no singularities for $x=0$, even if $\mathcal{D}(x)$ when $x=0$.
Since $\mathcal{D}(x)=C_0x+p(x)$, see Eq. (\ref{dcp}), the fraction appearing in Eq. (\ref{dave4}) can be rewritten as
{
\begin{eqnarray}
   \frac{2x\mathcal{D}(x)-x^2\mathcal{D}'(x)}{\mathcal{D}^2(x)}=\frac{C_0+2\frac{p(x)}{x}-p'(x)}{C_0^2+2C_0\frac{p(x)}{x}+\frac{p^2(x)}{x^2}}.
   \label{fraction}
\end{eqnarray}
}
Here, the function $p(x)$ is given in Eq. (\ref{pdx}) and, therefore, we obtain
\begin{align}
    \frac{p(x)}{x}=\sum_{\begin{subarray}{c}
			k=-\infty\\
			k\neq 0
		\end{subarray}}^{+\infty}\frac{C_k L}{2\pi i k}\frac{\exp{\left(\frac{2\pi i k x}{L}\right)}-1}{x}.
    \label{pdxsux}
\end{align}
We can use the well-known property stating that $(e^h-1)/h$ approaches 1 when $h$ approaches 0. 
So doing, we get the limiting behavior $p(x)/x\to\sum_{k=-\infty,\, k\neq 0}^{+\infty}C_k=1/g(0)-C_0$, when $x\to 0$. By using this result in Eq. (\ref{fraction}), we finally obtain the limiting behavior
{
\begin{eqnarray}
  \lim_{x\to 0} \frac{2x\mathcal{D}(x)-x^2\mathcal{D}'(x)}{\mathcal{D}^2(x)}=g(0),
   \label{fractionbis}
\end{eqnarray}
}
which excludes any possible singularity in the same fraction.
As before, when the time $t$ is large, the exponential in Eq. (\ref{dave4}) is increasingly flat and close to one, and the area for large values of $x$ becomes more and more important. Thus, being $p(x)$ bounded and periodic, we can neglect $p(x)/x$ and $p^2(x)/x^2$ in Eq. (\ref{fraction}) for large time, and we get
\begin{eqnarray}
\nonumber
    \overline{D}&=&\lim_{t\to \infty}\int_{-\infty}^{+\infty}\frac{C_0-p'(x)}{C_0^2}\frac{\exp\left(-\frac{C_0^2x^2}{4t}\right)}{\sqrt{4\pi t}}\mathrm{d} x.
    \label{dave5}
\end{eqnarray}
Recalling the Fourier development for $p(x)$ and applying the integral formula 
\begin{eqnarray}
\int_{-\infty}^{+\infty}e^{-ax^2}e^{ibx}dx=\sqrt{\frac{\pi}{a}}e^{-\frac{1}{4}\frac{b^2}{a}},
\label{int2}
\end{eqnarray}
we easily achieve the same result as obtained in Eq. (\ref{deff}).
{This new verification has the advantage of having used the integral in Eq. (\ref{int2}), which is simpler than Eq. (\ref{int1}).
 However, the function that is integrated to find the diffusion coefficient must be worked out properly, as seen above, in order to apply this simplification.}

{
\section{Explicit calculation of the quantity $\mathcal{D}(x)$ for the sinusoidal profile}

\label{appeB}

We calculate here the closed form expression for the quantity $\mathcal{D}(x)$, starting from Eq. (\ref{ddx}) of the main text, for the sinusoidal heterogeneous diffusion. 
We must then sum the series contained in that equation.}
To begin, we define $z=2\pi x/L$, and we observe that
\begin{align}
\nonumber
 \sum_{\begin{subarray}{c}
			k=-\infty\\
			k\neq 0
		\end{subarray}}^{+\infty}\frac{\beta^{\vert k \vert}}{k}e^{ikz}&=\sum_{k=1}^{+\infty}\frac{\left(\beta e^{iz}\right)^k}{k}-\sum_{k=1}^{+\infty}\frac{\left(\beta e^{-iz}\right)^k}{k}\\
		&=\mathcal{F}_\beta(z)-\mathcal{F}_\beta(-z),
\end{align}
where we introduced $\mathcal{F}_\beta(z)=\sum_{k=1}^{+\infty}{\left(\beta e^{iz}\right)^k}/{k}$.
In order to calculate a closed form expression for $\mathcal{F}_\beta(z)$, we start by observing that
\begin{align}
\sum_{k=1}^{+\infty}\left(\beta e^{iz}\right)^k=\frac{\beta e^{iz}}{1-\beta e^{iz}}=\frac{\beta(\cos z-\beta)+i\beta\sin z}{1-2\beta\cos z+\beta^2},
\label{seriesa}
\end{align}
where the geometric series is always convergent since $\beta^2<1$. If we integrate term by term the series in Eq. (\ref{seriesa}) we get
\begin{align}
\int_0^s\sum_{k=1}^{+\infty}\left(\beta e^{iz}\right)^k\mathrm{d} z=\sum_{k=1}^{+\infty}\frac{\beta^k}{ik} \left(e^{iks}-1\right),
\label{seriesb}
\end{align}
and therefore we can write
{
\begin{align}
\nonumber
    \mathcal{F}_\beta(s)&=\sum_{k=1}^{+\infty}\frac{\beta^k}{k}+i\mathcal{R}(s)-\mathcal{I}(s)\\&=\ln\frac{1}{1-\beta}+i\mathcal{R}(s)-\mathcal{I}(s),
    \label{fbeta}
\end{align}
}
where we used the classical logarithmic series and we defined the real and imaginary parts of the integral of Eq. (\ref{seriesa}), as follows 
\begin{align}
\mathcal{R}(s)&=\int_0^s\frac{\beta(\cos z-\beta)}{1-2\beta\cos z+\beta^2}\mathrm{d} z,\\
\mathcal{I}(s)&=\int_0^s\frac{\beta\sin z}{1-2\beta\cos z+\beta^2}\mathrm{d} z\, .
\end{align}
The first integral can be  tackled through the substitution $t=\tan(z/2)$, which leads to the new form
\begin{align}
\mathcal{R}(s)&=\frac{2\beta}{(1+\beta)^2}\int_0^{\tan\frac{s}{2}}\frac{1-t^2-\beta(1+t^2)}{\left[t^2+\left(\frac{1-\beta}{1+\beta}\right)^2\right](t^2+1)}\mathrm{d} t.
\end{align}
After partial fraction decomposition, we obtain
\begin{align}
\mathcal{R}(s)&=\int_0^{\tan\frac{s}{2}}\left[\frac{\frac{1-\beta}{1+\beta}}{t^2+\left(\frac{1-\beta}{1+\beta}\right)^2}-\frac{1}{t^2+1}\right]\mathrm{d} t,
\end{align}
and the final result is
\begin{align}
\mathcal{R}(s)=\arctan\left(\frac{1+\beta}{1-\beta}\tan\frac{s}{2}\right)-\arctan\left(\tan\frac{s}{2}\right).
\end{align}
The second integral turns elementary with the substitution $t=\cos z$, and the result follows
\begin{align}
\mathcal{I}(s)=\frac{1}{2}\ln\frac{1-2\beta\cos s+\beta^2}{1-2\beta+\beta^2}.
\end{align}
Summing up, thanks to Eq. (\ref{fbeta}), we can obtain the closed form for the function $\mathcal{F}_\beta(s)$
\begin{align}
\nonumber
    \mathcal{F}_\beta(s)=&\ln\frac{1}{1-\beta}+i\arctan\left(\frac{1+\beta}{1-\beta}\tan\frac{s}{2}\right)-i\arctan\left(\tan\frac{s}{2}\right)\\    
    \nonumber
    &-\frac{1}{2}\ln\frac{1-2\beta\cos s+\beta^2}{1-2\beta+\beta^2}\\
    \nonumber
    =&\frac{1}{2}\ln\frac{1}{1-2\beta\cos s+\beta^2}\\
    &+i\left[\arctan\left(\frac{1+\beta}{1-\beta}\tan\frac{s}{2}\right)-\arctan\left(\tan\frac{s}{2}\right)\right].
    \label{fbetafin}
\end{align}
Incidentally, by separating the real and imaginary parts of the series defining the function $\mathcal{F}_\beta(s)$, we obtain these two results
\begin{align}
\sum_{k=1}^{+\infty}\frac{\beta^k}{k}\cos(ks)&=\frac{1}{2}\ln\frac{1}{1-2\beta\cos s+\beta^2},\\
\sum_{k=1}^{+\infty}\frac{\beta^k}{k}\sin(ks)&=\arctan\left(\frac{1+\beta}{1-\beta}\tan\frac{s}{2}\right)-\arctan\left(\tan\frac{s}{2}\right),
\end{align}
{
which can be also proved with different standard techniques to sum complex series. }
Coming back to the heterogeneous sinusoidal diffusion, we can obtain the function $\mathcal{D}(x)$ through Eq. (\ref{ddx})
{
\begin{align}
    \mathcal{D}(x)=\frac{1}{G_0\sqrt{1-\varepsilon^2}}\left\lbrace x +\frac{L}{2\pi i} \left[\mathcal{F}_\beta\left(\frac{2\pi x}{L}\right)-\mathcal{F}_\beta\left(-\frac{2\pi x}{L}\right)\right]\right\rbrace.
		\label{ddx1}
\end{align}
}
By substituting Eq. (\ref{fbetafin}) in Eq. (\ref{ddx1}) and by considering the relationship $\frac{1+\beta}{1-\beta}=\sqrt{\frac{1-\varepsilon}{1+\varepsilon}}$, we get the final result
\begin{align}
\nonumber
    \mathcal{D}(x)=&\frac{1}{G_0\sqrt{1-\varepsilon^2}}\left\lbrace x +\frac{L}{\pi} \left[\arctan\left(\sqrt{\frac{1-\varepsilon}{1+\varepsilon}}\tan\frac{\pi x}{L}\right)\right.\right.\\
    &\left.\left.-\arctan\left(\tan\frac{\pi x}{L}\right)\right]\right\rbrace,
		\label{ddx2app}
\end{align}
which proves Eq. (\ref{ddx2}) of the main text.
In making the above substitution, we observed that the real part of $\mathcal{F}_\beta$ is an even function while the imaginary part is an odd function.

{
\section{Solution of the differential equation for $k(x)$}
\label{appeC}

We describe here the solution of the differential equation for the quantity $k(x)$, defined in Eq. (\ref{eqfork}).}
To solve Eq. (\ref{eqfork}), we first consider the left region $x_0-nL \le x < x_0$, and here we have
\begin{align}
    0=\frac{\mathrm{d}}{\mathrm{d} x}\left\lbrace\frac{1}{\mathcal{A}(x)}\frac{\mathrm{d} }{\mathrm{d} x}\left[\frac{k(x)}{\mathcal{B}(x)}\right]\right\rbrace,
    \end{align}
from which we can take
\begin{align}
    \frac{1}{\mathcal{A}(x)}\frac{\mathrm{d} }{\mathrm{d} x}\left[\frac{k(x)}{\mathcal{B}(x)}\right]=c_L,
    \end{align}
where $c_L$ is a constant ($L$ means left).
By integration we obtain
\begin{align}
     \int_{x_0-nL}^x\frac{\mathrm{d} }{\mathrm{d} \xi}\left[\frac{k(\xi)}{\mathcal{B}(\xi)}\right]\mathrm{d} \xi=c_L\int_{x_0-nL}^x\mathcal{A}(\xi)\mathrm{d} \xi,
\end{align}
or equivalently
\begin{align}
     \frac{k(x)}{\mathcal{B}(x)}-\frac{k(x_0-nL)}{\mathcal{B}(x_0-nL)}=c_L\int_{x_0-nL}^x\mathcal{A}(\xi)\mathrm{d} \xi.
\end{align}
Since $k(x_0-nL)=0$,  we get
\begin{align}
     k(x)=c_L\mathcal{B}(x)\int_{x_0-nL}^x\mathcal{A}(\xi)\mathrm{d} \xi,
     \label{leftsol}
\end{align}
in the left region $x_0-nL \le x < x_0$.

We consider now the right region $x_0<x\le x_0+nL$, and we have from Eq. (\ref{eqfork}) the equation
\begin{align}
    0=\frac{\mathrm{d}}{\mathrm{d} x}\left\lbrace\frac{1}{\mathcal{A}(x)}\frac{\mathrm{d} }{\mathrm{d} x}\left[\frac{k(x)}{\mathcal{B}(x)}\right]\right\rbrace,
    \end{align}
from which we find
\begin{align}
    \frac{1}{\mathcal{A}(x)}\frac{\mathrm{d} }{\mathrm{d} x}\left[\frac{k(x)}{\mathcal{B}(x)}\right]=c_R,
    \end{align}
where $c_R$ is another constant ($R$ means right).
Again by integration we get
\begin{align}
     \int_{x}^{x_0+nL}\frac{\mathrm{d} }{\mathrm{d} \xi}\left[\frac{k(\xi)}{\mathcal{B}(\xi)}\right]\mathrm{d} \xi=c_R\int_{x}^{x_0+nL}\mathcal{A}(\xi)\mathrm{d} \xi,
\end{align}
or equivalently
\begin{align}
     \frac{k(x_0+nL)}{\mathcal{B}(x_0+nL)}-\frac{k(x)}{\mathcal{B}(x)}=c_R\int_{x}^{x_0+nL}\mathcal{A}(\xi)\mathrm{d} \xi.
\end{align}
Since $k(x_0+nL)=0$,  we get
\begin{align}
     k(x)=-c_R\mathcal{B}(x)\int_{x}^{x_0+nL}\mathcal{A}(\xi)\mathrm{d} \xi,
     \label{rightsol}
\end{align}
in the right region $x_0 < x \le x_0+nL$.

We search now for the connection conditions for the two solutions in Eqs. (\ref{leftsol}) and (\ref{rightsol}). 
To this end, we integrate Eq. (\ref{eqfork}) in a small symmetric interval of radius $\varepsilon$ around $x_0$, namely
\begin{align}
    -\int_{x_0-\varepsilon}^{x_0+\varepsilon}\delta(x-x_0)\mathrm{d} x=\int_{x_0-\varepsilon}^{x_0+\varepsilon}\frac{\mathrm{d}}{\mathrm{d} x}\left\lbrace\frac{1}{\mathcal{A}(x)}\frac{\mathrm{d} }{\mathrm{d} x}\left[\frac{k(x)}{\mathcal{B}(x)}\right]\right\rbrace\mathrm{d} x,
    \label{eqforkint}
\end{align}
which yields
\begin{align}
    -1=\left.\frac{1}{\mathcal{A}(x)}\frac{\mathrm{d} }{\mathrm{d} x}\left[\frac{k(x)}{\mathcal{B}(x)}\right]\right|_{x_0-\varepsilon}^{x_0+\varepsilon}=c_R-c_L.
\end{align}
Moreover, we remember that $W(x,t)$ is a continuous function of $x$ for $x\in(x_0-nL,x_0+nL)$. In particular, $W(x_0^-,t)=W(x_0^+,t)$, and since $k(x)=\int_0^{+\infty}W(x,t)\mathrm{d} t$, we deduce that   $k(x_0^-)=k(x_0^+)$.
Given that the function $\mathcal{A}(x)$ is periodic, Eqs. (\ref{leftsol}) and (\ref{rightsol}), combined with $k(x_0^-)=k(x_0^+)$, result in $c_L=-c_R$.
Hence, the system composed by the two connection conditions $-1=c_R-c_L$ and $c_L=-c_R$ delivers 
$c_L = -c_R = 1/2$, and the final solution for $k(x)$ is
\begin{align}
     k(x)=\left\lbrace\begin{array}{cc}
          \frac{1}{2}\mathcal{B}(x)\int_{x_0-nL}^x\mathcal{A}(\xi)\mathrm{d} \xi, &  x_0-nL \le x < x_0, \\
          \frac{1}{2}\mathcal{B}(x)\int_{x}^{x_0+nL}\mathcal{A}(\xi)\mathrm{d} \xi, &  x_0 < x \le x_0+nL,
     \end{array}\right.
     \label{ksolapp}
\end{align}
which corresponds to Eq.(\ref{ksol}).

\section{Homogenization approach}
\label{appeD}
We consider Eq. (\ref{fptosolve}), and we prove that the obtained effective diffusion constant in Eq. (\ref{resres}) is coherent with the following ad-hoc homogenization procedure. In particular, we take into account the stationary version of Eq. (\ref{fptosolve}), which reads
\begin{align}
      0=\frac{\mathrm{d}}{\mathrm{d} x}\left\lbrace\frac{1}{\mathcal{A}(x)}\frac{\mathrm{d} }{\mathrm{d} x}\left[\frac{1}{\mathcal{B}(x)}W(x)\right]\right\rbrace,
\end{align}
and we introduce the two boundary conditions $W(0)=W_a$ and $W(nL)=W_b$, with $x\in(0,nL)$.
{
As before, we use the differential operator $\frac{\mathrm{d}}{\mathrm{d} x}$ instead of $\frac{\partial}{\partial x}$ since the time variable is no longer present.}
The steady-state condition involves a constant flow $J$, given by 
\begin{align}
    \frac{1}{\mathcal{A}(x)}\frac{\mathrm{d} }{\mathrm{d} x}\left[\frac{1}{\mathcal{B}(x)}W(x)\right]=-J.
    \label{flux}
\end{align}
When $\mathcal{A}(x)$ and $\mathcal{B}(x)$ are constant functions, we find that $J=-\frac{1}{\mathcal{A}\mathcal{B}}\frac{\mathrm{d} W(x)}{\mathrm{d} x}$, and therefore the effective diffusion constant can be written as
\begin{align}
    D_{\it eff} = \frac{1}{\mathcal{A}\mathcal{B}}=-\frac{JnL}{W_b-W_a},
    \label{dconst}
\end{align}
since in this case the gradient is obtained as  $\frac{\mathrm{d} W(x)}{\mathrm{d} x}=(W_b-W_a)/(nL)$.

In the general case with $\mathcal{A}(x)$ and $\mathcal{B}(x)$ being periodic functions, we can integrate Eq. (\ref{flux}), eventually obtaining 
\begin{align}
    \frac{W(x)}{\mathcal{B}(x)}-\frac{W(0)}{\mathcal{B}(0)}=-J\int_0^x\mathcal{A}(\xi)\mathrm{d}\xi,
\end{align}
or equivalently
\begin{align}
    W(x)=\mathcal{B}(x)\left[\frac{W(0)}{\mathcal{B}(0)}-J\int_0^x\mathcal{A}(\xi)\mathrm{d}\xi\right].
\end{align}
This is the behavior of the density over the considered interval, that is for $x\in(0,nL)$. From this expression, we can also obtain a similar result for $W(x+L)$, which assumes the form
\begin{align}
    W(x+L)=\mathcal{B}(x+L)\left[\frac{W(0)}{\mathcal{B}(0)}-J\int_0^{x+L}\mathcal{A}(\xi)\mathrm{d}\xi\right].
\end{align}
By using the periodicity of $\mathcal{B}(x)$, we obtain
\begin{align}
    W(x+L)-W(x)=-J\mathcal{B}(x)\int_x^{x+L}\mathcal{A}(\xi)\mathrm{d}\xi,
\end{align}
and by using the periodicity of $\mathcal{A}(x)$, we can write
\begin{align}
    W(x+L)-W(x)=-J\mathcal{B}(x)\int_0^{L}\mathcal{A}(\xi)\mathrm{d}\xi.
\end{align}
It means that the function $ W(x+L)-W(x)$ is periodic and it is proportional to $\mathcal{B}(x)$.
We can calculate the average value over one period of the function $ W(x+L)-W(x)$ and we get
\begin{align}
    \left\langle W(x+L)-W(x) \right\rangle=-\frac{J}{L}\int_0^{L}\mathcal{B}(\eta)\mathrm{d}\eta\int_0^{L}\mathcal{A}(\xi)\mathrm{d}\xi.
    \label{aveave}
\end{align}
Similarly to the approach discussed for finding Eq. (\ref{dconst}), we can assume that 
\begin{align}
    D_{\it eff} =-\frac{JL}{\left\langle W(x+L)-W(x) \right\rangle}.
    \label{effeff}
\end{align}
This relationship corresponds to assuming that $W_b-W_a=N\left\langle W(x+L)-W(x) \right\rangle$, as can be seen immediately from Eq. (\ref{dconst}). This turns out to be quite reasonable when we consider that the density jump between 0 and $nL$ is given by the sum of the density jumps over all the periods composing the total interval and that the function $ W(x+L)-W(x)$ is periodic.
Anyway, by combining Eqs. (\ref{aveave}) and (\ref{effeff}), we immediately obtain the result
\begin{align}
D_{\it eff} = \frac{L^2}{\int_{0}^{L}\mathcal{B}(\eta)\mathrm{d} \eta\int_{0}^{L}\mathcal{A}(\xi)\mathrm{d} \xi}=\frac{1}{\left\langle \mathcal{A} \right\rangle\left\langle \mathcal{B} \right\rangle},
\end{align}
which is consistent with Eq. (\ref{resres}).

\begin{acknowledgments}

S. G. and R. B. acknowledge
support funding of the French National Research Agency ANR through project `Dyprosome' (ANR-21-CE45-0032-02). 
\end{acknowledgments}

\bibliography{fokkerplanck}

\end{document}